\documentclass[american,aps,pra,reprint,superscriptaddress,twocolumn,showpacs]{revtex4-1} 
\usepackage[unicode=true,pdfusetitle, bookmarks=true,bookmarksnumbered=false,bookmarksopen=false, breaklinks=false,pdfborder={0 0 0},backref=false,colorlinks=false] {hyperref}
\hypersetup{ colorlinks,linkcolor=myurlcolor,citecolor=myurlcolor,urlcolor=myurlcolor}
\usepackage{braket,colortbl,cleveref,amsthm,amsmath,amssymb,txfonts}
\definecolor{myurlcolor}{rgb}{0,0,0.7}
\usepackage{graphics,graphicx}
\usepackage{color}
\usepackage{colortbl}
\usepackage{subfigure}
\usepackage{graphicx}
\usepackage[format = plain,labelfont = bf,up, textfont = normal , up, justification =raggedright, singlelinecheck =false]{caption}

\usepackage{amsmath}
\usepackage{graphicx}
\usepackage[utf8x]{inputenc}
\usepackage{color}
\usepackage{amsmath}
\usepackage{braket}
\usepackage{tikz}
\usetikzlibrary{shapes.geometric, arrows}
\usepackage{latexsym}
\usepackage{amssymb}
\usepackage{amsthm}
\usepackage{bm}
\usepackage{graphics,epstopdf}
\usepackage{color}\usepackage{amsmath}

\usepackage[usenames,dvipsnames,svgnames]{pstricks}
\usepackage{epsfig}
\usepackage{pst-grad} 
\tikzstyle{startstop} = [rectangle, rounded corners, minimum width=3cm, minimum height=1cm,text centered, draw=black, fill=red!30]

\tikzstyle{env}=[circle,  ball color = green!20, minimum size= 80mm]
\tikzstyle{central}=[circle, ball color = red!100, minimum size=8mm]
\tikzstyle{bath}=[circle, ball color =blue!75, minimum size=4mm]

\theoremstyle{plain}

\def\bea{\begin{eqnarray}}
\def\eea{\end{eqnarray}}
\def\ba{\begin{array}}
\def\ea{\end{array}}

\def\ket{\rangle}
\def\bra{\langle}
\def\beq{\begin{equation}}
\def\eeq{\end{equation}}

\begin{document}

\title{Dynamics and thermodynamics of a central spin immmersed in a spin bath}
\author{Chiranjib Mukhopadhyay}
\email{chiranjibmukhopadhyay@hri.res.in}
\affiliation{Harish-Chandra Research Institute, Allahabad 211019, India  and\\
 Homi Bhabha National Institute, Training School Complex, Anushakti Nagar, Mumbai 400 085, India.}
 
\author{Samyadeb Bhattacharya}
\email{samyadebbhattacharya@hri.res.in}
\affiliation{Harish-Chandra Research Institute, Allahabad 211019, India  and\\
 Homi Bhabha National Institute, Training School Complex, Anushakti Nagar, Mumbai 400 085, India.}
\author{Avijit Misra}
\email{avijitm@imsc.res.in}
 \affiliation{Optics and Quantum Information Group, The Institute of Mathematical Sciences,\\H.B.N.I., C.I.T. campus, Taramani, Chennai 600113, India.}

\author{Arun Kumar Pati}
\email{akpati@hri.res.in}
\affiliation{Harish-Chandra Research Institute, Allahabad 211019, India  and\\
 Homi Bhabha National Institute, Training School Complex, Anushakti Nagar, Mumbai 400 085, India.}



\begin{abstract}
 An exact reduced dynamical map along with its operator sum representation is derived for a central spin interacting with a thermal spin environment. The dynamics of the central spin shows high sustainability of quantum
traits like coherence and entanglement in the low temperature \textcolor{black}{regime}. However, for sufficiently high temperature and
when the number of bath particles approaches the thermodynamic limit, this feature vanishes and the dynamics
closely mimics  Markovian evolution. The properties
of the long time averaged state and the trapped information of the initial state for the central qubit are also investigated in detail, confirming that the non-ergodicity of the dynamics can be attributed to the finite temperature and finite size of the bath. It is shown that if a certain stringent resonance condition is satisfied, the long time averaged state retains quantum coherence, which can have far reaching technological implications in engineering quantum devices. An exact time local master equation of the canonical form is derived . With the help of this master equation, the non-equilibrium properties of the central spin system are studied by investigating the detailed balance condition and irreversible entropy production rate. The result reveals that the central qubit thermalizes only in the
limit of very high temperature and large number of bath spins.


\end{abstract}

\maketitle

\section{Introduction}
\label{I}
In the microscopic world, physical systems are rarely isolated from environmental influence. Systems relevant for implementation of quantum information theoretic and computational 
tasks like ion traps \cite{ciraczoller},quantum dots \cite{dots}, NMR qubits \cite{nmr}, polarized photons \cite{klm}, Josephson junction qubits \cite{squid} or NV centres \cite{nvcenter, nvreview} 
all interact with their respective environments to some extent. Therefore it is necessary to study the properties of open system dynamics for such quantum systems immersed in baths. For quantum systems
exposed to usual Markovian baths, their quantumness gradually fades over time, thus negating any advantage gained through the use of quantum protocols over classical ones. Even in thermodynamics, the presence
of quantum coherence \citep{brask,huber} or entanglement \citep{brunnerent} enhances the performance of quantum heat machines. Thus, it is imperative to engineer baths in such a way so as to retain nonclassical 
features of the system for large durations.

\begin{figure}

\begin{center}
\begin{tikzpicture}[node distance=3cm]
\node (env)[env, xshift = 0 cm]{} ;
\node (central)[central, xshift = 0 cm]{} ;
\foreach \a in {1,2,...,25}{
\draw (\a*360/25: 2.7 cm) node (bathspin \a) [bath]{};
\draw [thick, dotted] (central) --(bathspin \a);
;}
\end{tikzpicture}
\end{center}
\caption{(Colour online) Schematic diagram of the central spin model. The central spin (red circle) interacts with the bath (green) constituting of spins (blue circles).}
\label{modelschematic}
\end{figure}
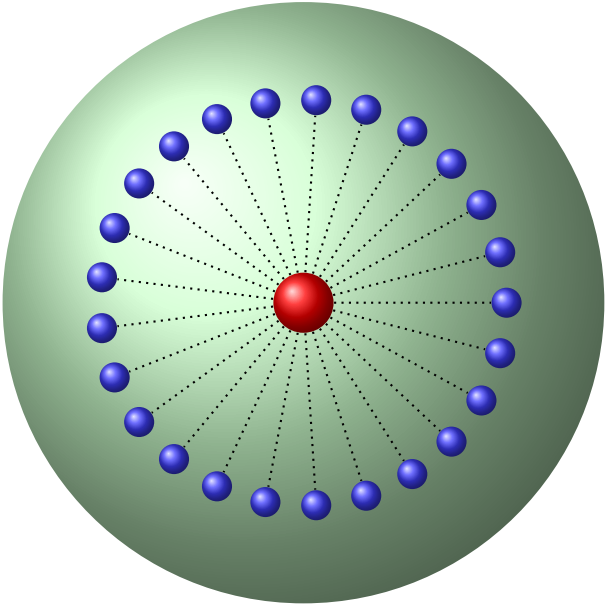

Baths can be broadly classified into two different classes, namely Bosonic and Fermionic. Paradigmatic examples for Bosonic baths include the Caldeira-Leggett model \cite{caldeiraleggett} or the spin Boson model \cite{rmpbosonic}.
Lindblad type master equations for these models can be found in the literature \cite{breuerbook}. However, in the Fermionic case, where one models the bath as a collection of a large number of spin-$\frac{1}{2}$ particles, 
the situation is generally trickier and one often has to rely on perturbative techniques or time nonlocal master equations \cite{spinstar1,spinstar2}. Far from from being a theoretical curiosity, the solution of such systems
is of paramount importance in physical situations such as magnetic systems \citep{magnets}, quantum spin glasses \citep{spinglass} or superconducting systems \citep{conductor}. \\
One specific example of a qubit immersed in a  Fermionic bath is the Non-Markovian spin star model (schematic diagram in Fig. \ref{modelschematic}) \cite{spinstar1,spinstar2,spinstar3,sagnik}, which 
is relevant for quantum computing with NV centre \cite{nvcomp} defects within a diamond lattice. We show that it is possible to preserve coherence and entanglement in this system for quite a long time by
choosing bath parameter values appropriately. Even more interestingly we confirm the presence of quantum coherence in the system for the long time averaged state for certain resonance conditions, which is
an utter impossibility for the usual Markovian thermal baths. Such strict and fragile resonance conditions underlie our emphasis on the need for ultra-precise engineering of the bath. We also investigate the 
amount of information trapped \citep{smirne} in the central spin system and draw a connection of the same with the process of equilibration. \\
A time nonlocal integrodifferential master equation was set up for the central spin model using the correlated projection operator technique in Ref. \citep{spinstar2}.
An exact time local master equation for this system was derived in the limit of infinite bath temperature in Ref. \cite{samyadeb} from the corresponding reduced dynamical map. 
In this paper, we considerably extend the scope of previous results by deriving the exact reduced dynamics and the exact Lindblad type master equation for arbitrary bath temperature and
system bath coupling strength. Our formalism allows us to study the approach towards equilibration in sufficient detail. 

The  paper is organised as follows. In Section \ref{II} we introduce the central spin model and find the exact reduced dynamics for the system and the corresponding Kraus operator representation. We use the solution for the exact reduced dynamics to study the evolution of quantum coherence and entanglement. In Section \ref{IV} we study the long time averaged state and its properties. We analyse the resonance condition for the existence of quantum coherence even in the long time averaged state and the phenomenon of information trapping in the central qubit. In Section \ref{III}, we begin with the derivation of the exact time-local master equation for this system and use this master equation to investigate the non-equilibrium nature of the dynamics through a thorough study of the deviation from the detailed balance condition as well as the temporal dependence of irreversible entropy production rate. We finally conclude in Section \ref{V}. 

\section{Central spin model and its reduced dynamics}
\label{II}
In this section we present the model for the qubit coupled centrally to a thermal spin bath. Then we derive the exact dynamical map for the qubit. We also derive the Kraus operators for the reduced dynamics.
\subsection{The model}
We consider a spin-$\frac{1}{2}$ particle interacting uniformly with $N$ other mutually non-interacting spin-$\frac{1}{2}$ particles constituting the bath. 

\noindent The total Hamiltonian for this spin bath model is given by 
\begin{eqnarray}
 \label{sec1a}
H&=&H_S+H_B+H_{SB}\\ \nonumber
 &=&\frac{\hbar}{2}\omega_0\sigma_z^0+\frac{\hbar\omega}{2N}\sum_{i=1}^{N}\sigma_z^i+\frac{\hbar\epsilon}{2\sqrt{N}}\sum_{i=1}^N (\sigma_x^0\sigma_x^i+\sigma_y^0\sigma_y^i),
\end{eqnarray}
with $\sigma_k^i$ ($k=x,y,z$) as the Pauli matrices of the i-th spin of the bath and $\sigma_k^0$ ($k=x,y,z$) as the same for the central spin and $\epsilon$ is the system-bath interaction parameter. Here $H_S$, $H_B$ and $H_{SB}$ are the system, bath and interaction Hamiltonian respectively. $N$ is the number of bath atoms directly interacting with the central spin. The bath frequency and the system-bath interaction strength are both rescaled as $\omega/N$ and $\epsilon/\sqrt{N}$ respectively. By the use of collective angular momentum operators for the bath spins $J_{l}=\sum_{i=1}^N \sigma_l^i$  (where $l=x,y,z,+,-$), we rewrite the bath and interaction Hamiltonians as 
\beq\label{sec1b}
\begin{array}{ll}
H_B=\frac{\hbar\omega}{2N}J_z,\\
H_{SB}=\frac{\hbar\epsilon}{2\sqrt{N}}(\sigma_x^0J_x+\sigma_y^0J_y).
\end{array}
\eeq
We then use the Holstein-Primakoff transformation \cite{holsteinprimakoff,prbchinese} to redefine the collective bath angular momentum operators as
\beq\label{sec1c}
J_+=\sqrt{N}b^{\dagger}\left(1-\frac{b^{\dagger}b}{2N}\right)^{1/2}~~,~~J_-=\sqrt{N}\left(1-\frac{b^{\dagger}b}{2N}\right)^{1/2}b,
\eeq
where $b$ and $b^{\dagger}$ are the bosonic annihilation and creation operators with the property $[b,b^{\dagger}]=1$. Then the Hamiltonians of Eq. (\ref{sec1b}) can be rewritten as 
\beq\label{sec1d}
\begin{array}{ll}
H_B~~=-\frac{\hbar\omega}{2}\left(1-\frac{b^{\dagger}b}{N}\right),\\
H_{SB}=\hbar\epsilon\left[\sigma_0^+\left(1-\frac{b^{\dagger}b}{2N}\right)^{1/2}b+\sigma_0^-b^{\dagger}\left(1-\frac{b^{\dagger}b}{2N}\right)^{1/2}\right].
\end{array}
\eeq
\subsection{Dynamical map of the central spin}
In the following, we derive the exact reduced dynamical map of the central spin after performing the Schr\"{o}dinger evolution for the total system and bath and then tracing over the bath degrees of freedom. It is assumed that the initial system bath joint state is a product state $\rho_{SB}(0)=\rho_S(0)\otimes\rho_B(0)$, which ensures the complete positivity of the reduced dynamics \citep{gorini,lindblad}. The initial bath state is considered as a thermal state $\rho_B(0)=e^{-H_B/KT}/Z$, where $K$, $T$ and $Z$ are the Boltzman constant, temperature of the bath and the partition function respectively. Consider the evolution of the state $|\psi(0)\ket=|1\ket|x\ket$, where $|1\ket$ is the system excited state and $|x\ket$ is an arbitrary bath state. After the unitary evolution $U(t)=\exp \left(-\frac{iHt}{\hbar}\right)$, let the state is $|\psi(t)\ket = \gamma_1(t)|1\ket|x'\ket +\gamma_2(t)|0\ket|x''\ket$. let us now define two operators $\hat{A}(t)$ and $\hat{B}(t)$ corresponding to the bath Hilbert space such that $\hat{A}(t)|x\ket=\gamma_1(t)|x'\ket$ and $\hat{B}(t)|x\ket=\gamma_2(t)|x''\ket$. Then we have $|\psi(t)\ket = \hat{A}(t)|1\ket|x'\ket +\hat{B}(t)|0\ket|x''\ket$. Now from the Schr\"{o}dinger equation $\frac{d}{dt}|\psi(t)\ket=-\frac{i}{\hbar}H|\psi(t)\ket$, we have 
\beq\label{a1}
\begin{array}{ll}
\frac{d\hat{A}(t)}{dt}=-i\left(\frac{\omega_0}{2}-\omega\left(1-\frac{b^{\dagger}b}{2N}\right)\right)\hat{A}(t)-i\epsilon\left(1-\frac{b^{\dagger}b}{2N}\right)^{1/2}b\hat{B}(t),\\
\frac{d\hat{B}(t)}{dt}=i\left(\frac{\omega_0}{2}+\omega\left(1-\frac{b^{\dagger}b}{2N}\right)\right)\hat{B}(t)-i\epsilon b^{\dagger}\left(1-\frac{b^{\dagger}b}{2N}\right)^{1/2}\hat{A}(t).
\end{array}
\eeq
By substituting $\hat{A}(t)=\hat{A}_1(t)$ and $\hat{B}(t)=b^{\dagger}\hat{B}_1(t)$, we have 
\beq\label{a2}
\begin{array}{ll}
\frac{d\hat{A}_1(t)}{dt}=-i\left(\frac{\omega_0}{2}-\omega\left(1-\frac{\hat{n}}{2N}\right)\right)\hat{A}_1(t)-i\epsilon\left(1-\frac{\hat{n}}{2N}\right)^{1/2}(\hat{n}+1)\hat{B}_1(t),\\
\frac{d\hat{B}_1(t)}{dt}=i\left(\frac{\omega_0}{2}+\omega\left(1-\frac{\hat{n}+1}{2N}\right)\right)\hat{B}_1(t)-i\epsilon\left(1-\frac{\hat{n}}{2N}\right)^{1/2}\hat{A}_1(t),
\end{array}
\eeq
where $\hat{n}=b^{\dagger}b$ is the number operator. The operator equations \eqref{a2} can be straight forwardly solved and the solutions will be functions of $\hat{n}$ and $t$. Then $\hat{A}_1(t)|n\ket=A_1(n,t)|n\ket$, where $\hat{n}|n\ket=n|n\ket$. Therefore the evolution of the reduced state of the qubit ($|1\ket\bra 1|$) can now be found by tracing over the bath modes as
\beq\label{a3}
\begin{array}{ll}
\phi(|1\ket\bra 1|)=Tr_B\left[|\psi(t\ket\bra\psi(t)|)\right]=\\
\frac{1}{Z}\sum_{n=0}^N \left(|A_1(n,t)|^2|1\ket\bra 1|+(n+1)|B_1(n,t)|^2|0\ket\bra 0|\right)e^{-\frac{\hbar\omega}{KT}(n/2N-1/2)},
\end{array}
\eeq
where from the solution of \eqref{a2}, we have $|B_1(n,t)|^2=4\epsilon^2(1-n/2N)\frac{\sin^2(\eta t/2)}{\eta}$ and $|A_1(n,t)|^2=1-(n+1)|B_1(n,t)|^2$.

Similarly we define $\chi(0)=|0\ket|x\ket$ and $\chi(t)=\hat{C}(t)|0\ket|x\ket+\hat{D}(t)|1\ket|x\ket$. Following the similar procedure and with the substitution $\hat{C}(t)=\hat{C}_1(t),~~\hat{D}(t)=b\hat{D}_1(t)$, we find
\beq\label{a4}
\begin{array}{ll}
\frac{d\hat{C}_1(t)}{dt}=i\left(\frac{\omega_0}{2}+\omega\left(1-\frac{\hat{n}}{2N}\right)\right)\hat{C}_1(t)-i\epsilon\hat{n}\left(1-\frac{\hat{n}-1}{2N}\right)^{1/2}\hat{D}_1(t),\\
\frac{d\hat{D}_1(t)}{dt}=-i\left(\frac{\omega_0}{2}-\omega\left(1-\frac{\hat{n}-1}{2N}\right)\right)\hat{D}_1(t)-i\epsilon\left(1-\frac{\hat{n}-1}{2N}\right)^{1/2}\hat{C}_1(t),
\end{array}
\eeq 
From the solution of \eqref{a4}, we find 
\beq\label{a5}
\begin{array}{ll}
\phi(|0\ket\bra 0|)=Tr_B\left[|\chi(t\ket\bra\chi(t)|)\right]\\
=\frac{1}{Z}\sum_{n=0}^N \left(n|D_1(n,t)|^2|1\ket\bra 1|+|C_1(n,t)|^2|0\ket\bra 0|\right)e^{-\frac{\hbar\omega}{KT}(n/2N-1/2)},
\end{array}
\eeq
with $|D_1(n,t)|^2=4\epsilon^2(1-(n-1)/2N)\frac{\sin^2(\eta' t/2)}{\eta'}$ and $|C_1(n,t)|^2=1-n|D_1(n,t)|^2$. For the off-diagonal component of the reduced density matrix, we have 
\beq\label{a6}
\begin{array}{ll}
\phi(|1\ket\bra 0|)=Tr_B\left[|\psi(t\ket\bra\chi(t)|)\right]\\
=\frac{1}{Z}\sum_{n=0}^N \left(A_1(n,t)C_1^*(n,t)|1\ket\bra 0|\right)e^{-\frac{\hbar\omega}{KT}(n/2N-1/2)},
\end{array}
\eeq
with $A_1(n,t)C_1^*(n,t)=\Delta(t)$. \\
Therefore the reduced state of the system after the unitary evolution of the joint system-bath state, can be expressed as 
\beq\label{sec1e}
\begin{array}{ll}
\rho_S(t)=\mbox{Tr}_B\left[e^{-iHt/\hbar}\rho_S(0)\otimes\rho_B(0)e^{iHt/\hbar}\right],\\
~~~~~~~~=\left(\begin{matrix}
\rho_{11}(t) & \rho_{12}(t)\\
\rho_{21}(t) & \rho_{22}(t)
\end{matrix}\right),
\end{array}
\eeq
where the components of the density matrix are given by 
\beq\label{sec1f}
\begin{array}{ll}
\rho_{11}(t)=\rho_{11}(0)(1-\alpha(t))+\rho_{22}(0)\beta(t),\\
\rho_{12}(t)=\rho_{12}(0)\Delta(t),
\end{array}
\eeq
with
\beq\label{sec1g}
\begin{array}{ll}
\alpha(t)=\frac{1}{Z}\sum_{n=0}^N 4(n+1)\epsilon^2\left(1-\frac{n}{2N}\right)\frac{\sin^2(\eta t/2)}{\eta^2}e^{-\frac{\hbar\omega}{KT}(n/2N-1/2)},\\
\\
\beta(t)=\frac{1}{Z}\sum_{n=0}^N 4n\epsilon^2\left(1-\frac{n-1}{2N}\right)\frac{\sin^2(\eta' t/2)}{\eta'^2}e^{-\frac{\hbar\omega}{KT}(n/2N-1/2)},\\
\\
\Delta(t)=\frac{1}{Z}\sum_{n=0}^N e^{-i\omega t/2N}\left(\cos(\eta t/2)-i(\omega_0-\omega/2N)\sin(\eta t/2)\right)\times\\
~~~~~~~~~~~~\left(\cos(\eta' t/2)+i(\omega_0-\omega/2N)\sin(\eta' t/2)\right)e^{-\frac{\hbar\omega}{KT}(n/2N-1/2)},
\end{array}
\eeq
and
\beq\label{sec1g1}
\begin{array}{ll}
\eta = \sqrt{\left(\omega_0-\frac{\omega}{2N}\right)^2+4\epsilon^2(n+1)\left(1-\frac{n}{2N}\right)},\\
\\
\eta' = \sqrt{\left(\omega_0-\frac{\omega}{2N}\right)^2+4\epsilon^2 n\left(1-\frac{n-1}{2N}\right)},
\end{array}
\eeq
where the partition function is $Z=\sum_{n=0}^N e^{-\frac{\hbar\omega}{KT}(n/2N-1/2)}$. 
\subsection{Operator sum representation}
A very important aspect of general quantum evolution, represented by completely positive trace preserving operation is the Kraus operator sum representation, given as $\rho(t)=\sum_i K_i(t)\rho(0)K_i^{\dagger}(t)$. The Kraus operators can be constructed \cite{leung} from the eigenvalues and eigenvectors of the corresponding Choi-Jamiolkowski (CJ) state \cite{choicp, leung}.
The CJ state for a dynamical map $\Phi[\rho]$ acting on a $d$ dimensional system is given by $(\mathbb{I}_d \otimes\Phi)[\Phi_{+}]$, with $\Phi_{+}=|\Phi_{+}\rangle\langle\Phi_{+}|$ being the maximally entangled state in $d\times d$ dimension. For the particular evolution considered here, we find the CJ state to be 
\beq\label{sec1newb}
\left(
\begin{matrix}
\frac{1-\alpha(t)}{2} && 0 && 0 && \frac{\Delta(t)}{2}\\
0 && \frac{\alpha(t)}{2} && 0 && 0\\
0 && 0 && \frac{\beta(t)}{2} && 0\\
\frac{\Delta^*(t)}{2} && 0 && 0 && \frac{1-\beta(t)}{2}
\end{matrix}
\right).
\eeq
From the eigensystem of the CJ state given in (\ref{sec1newb}), we derive the Kraus operators as  
\beq\label{sec1newa}
\begin{array}{ll}
K_1(t)=\sqrt{\beta(t)}\left(\begin{matrix}
                        0 && 1\\
                        0 && 0
                        \end{matrix}\right),\\
\\
K_2(t)=\sqrt{\alpha(t)}\left(\begin{matrix}
                        0 && 0\\
                        1 && 0
                        \end{matrix}\right),\\
\\
K_3(t)=\sqrt{\frac{X_1}{1+Y_1^2}}\left(\begin{matrix}
                        Y_1 e^{i\theta(t)} && 0\\
                        0 && 1
                        \end{matrix}\right),\\ 
\\
K_4(t)=\sqrt{\frac{X_2}{1+Y_2^2}}\left(\begin{matrix}
                        Y_2 e^{i\theta(t)} && 0\\
                        0 && 1
                        \end{matrix}\right),\\                      
\end{array}
\eeq
where $\theta(t)=\arctan[\Delta_I(t)/\Delta_R(t)]$ and 
$$ X_{1,2}=\left(1-\frac{\alpha(t)+\beta(t)}{2}\right)\pm \frac{1}{2}\sqrt{(\alpha(t)-\beta(t))^2+4|\Delta(t)|^2}, $$
$$Y_{1,2}=\frac{\sqrt{(\alpha(t)-\beta(t))^2+4|\Delta(t)|^2}\mp(\alpha(t)-\beta(t))}{2|\Delta(t)|}.$$

\noindent One can check that the Kraus operators satisfy the condition $\sum_i K_i^{\dagger}(t)K_i(t)=\mathbb{I}$.
\subsection{Coherence and Entanglement dynamics of the central spin}
\label{coh_subsection}
Having obtained the exact reduced dynamics of the central spin, in the following we study the temporal variation of non-classical properties, viz. quantum coherence and entanglement of the system. It is well known that for usual Markovian systems, such non-classical quantities decay monotonically over time and eventually disappear \citep{baumgratz,titas,yueberly}. However, the central spin system is strongly non-Markovian in nature and therefore, a natural and pertinent question is to ask whether it is possible to preserve quantum features for long periods of time for this system. The following subsections are devoted to answering that question for various parameter regimes of the spin bath model.

\textbf{Quantum Coherence}: In this article we consider \textbf{$l_{1}$}-norm of coherence as a quantifier of quantum coherence. For a qubit system, the \textbf{$l_{1}$}-norm of coherence \cite{baumgratz} $C_{l_{1}}$ is simply given by twice the absolute value of any off-diagonal element, i.e., $2 \vert \rho_{12}(t) \vert $. The evolution of coherence is then given by
\begin{equation}
\label{coherence1}
C_{l_{1}}(t) = C_{l_{1}}(0) |\Delta(t)|.
\end{equation}
This is a straightforward scaling of the initial quantum coherence. One immediate consequence is that we cannot create coherence over and above the coherence present in the system initially, even though this is a strongly non-Markovian system. In subsequent analysis, we can thus take the initial coherence to be unity, i.e. the maximally coherent state without loss of generality. 

\textbf{Quantum Entanglement}: Operationally, quantum entanglement is the most useful resource in quantum information theory \citep{e91, shor, densecoding, teleportation, horodeckireview}. However, it is also a fragile one \citep{suddendeath} and decays quite quickly for Markovian evolution \citep{yueberly}. We suppose a scenario in which the central spin qubit is initially entangled to an ancilla qubit $A$ in addition to the spin bath. There is no subsequent interaction between the ancilla qubit and the central spin. Our goal is to investigate the entanglement dynamics of the joint two-qubit state $\rho_{SA}$. From the factorization theorem for quantum entanglement \cite{factorization}, we have
\begin{equation}
\label{ent_fact}
E(\rho_{SA} (t)) = E(\rho_{SA} (0))E\left(\chi_{SA}(t)\right),
\end{equation}
where $\chi_{SA}(t)$ is the CJ State in \eqref{sec1newb} and the entanglement measure $E$ is  concurrence \cite{wootters}. Concurrence of a two qubit system is given as
$E(\rho_{AB}) = \max\{0,\lambda_1-\lambda_2-\lambda_3-\lambda_4\},$
where $\lambda_1,\dots,\lambda_4$ are the square roots of the eigenvalues of $\rho_{AB}\tilde{\rho}_{AB}$ in 
decreasing order, $\tilde{\rho}_{AB}= (\sigma_y \otimes \sigma_y)\rho_{AB}^{*}(\sigma_y \otimes \sigma_y)$. 
Here  the complex conjugation $\rho_{AB}^{*}$ is taken in the computational basis, and $\sigma_y$ is the Pauli spin matrix. From now on, we mean concurrence by entanglement throughout the paper. Then the entanglement of the CJ state can be written as  $E (\chi_{SA} (t))= max \left( 0, \vert \Delta (t) \vert - \sqrt{\alpha(t) \beta(t)}\right)$. Since the initial entanglement $ E(\rho_{SA} (0))$ is simply a constant scaling term, we take this to be unity, i.e. consider a maximally entangled initial $\rho_{SA} (0) $ state without loss of generality and study the subsequent dynamics.

\begin{widetext}

\begin{figure}[ht]
\centering
\subfigure[ Variation of coherence $C(t)$ with time $t$ for different bath temperature.]{
    \includegraphics[width=0.3\textwidth, keepaspectratio]{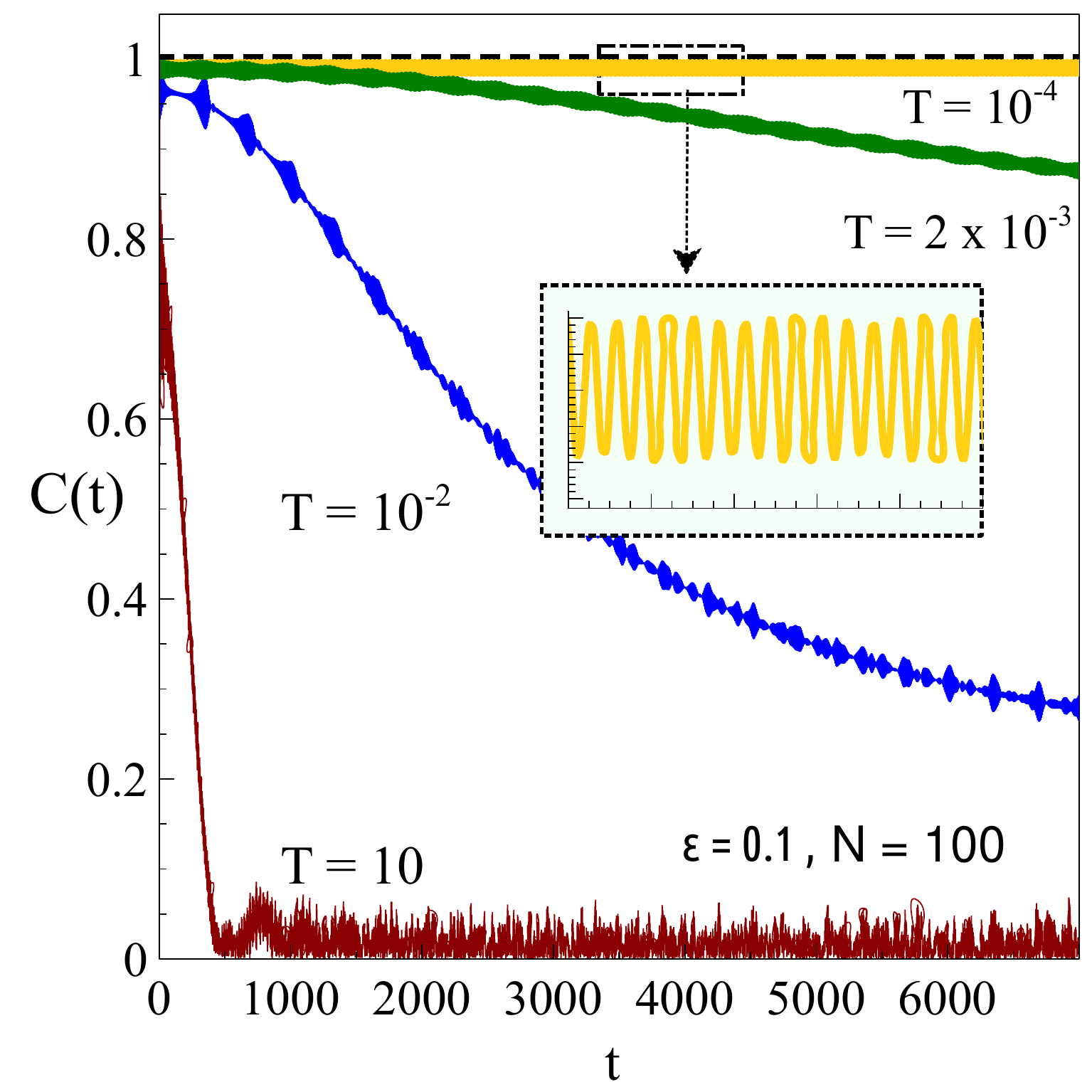}
    \label{coh-fig1}
}
\subfigure[Variation of coherence $C(t)$ with time $t$ for different system-bath interaction strength.]{
    \includegraphics[width=0.3\textwidth, keepaspectratio]{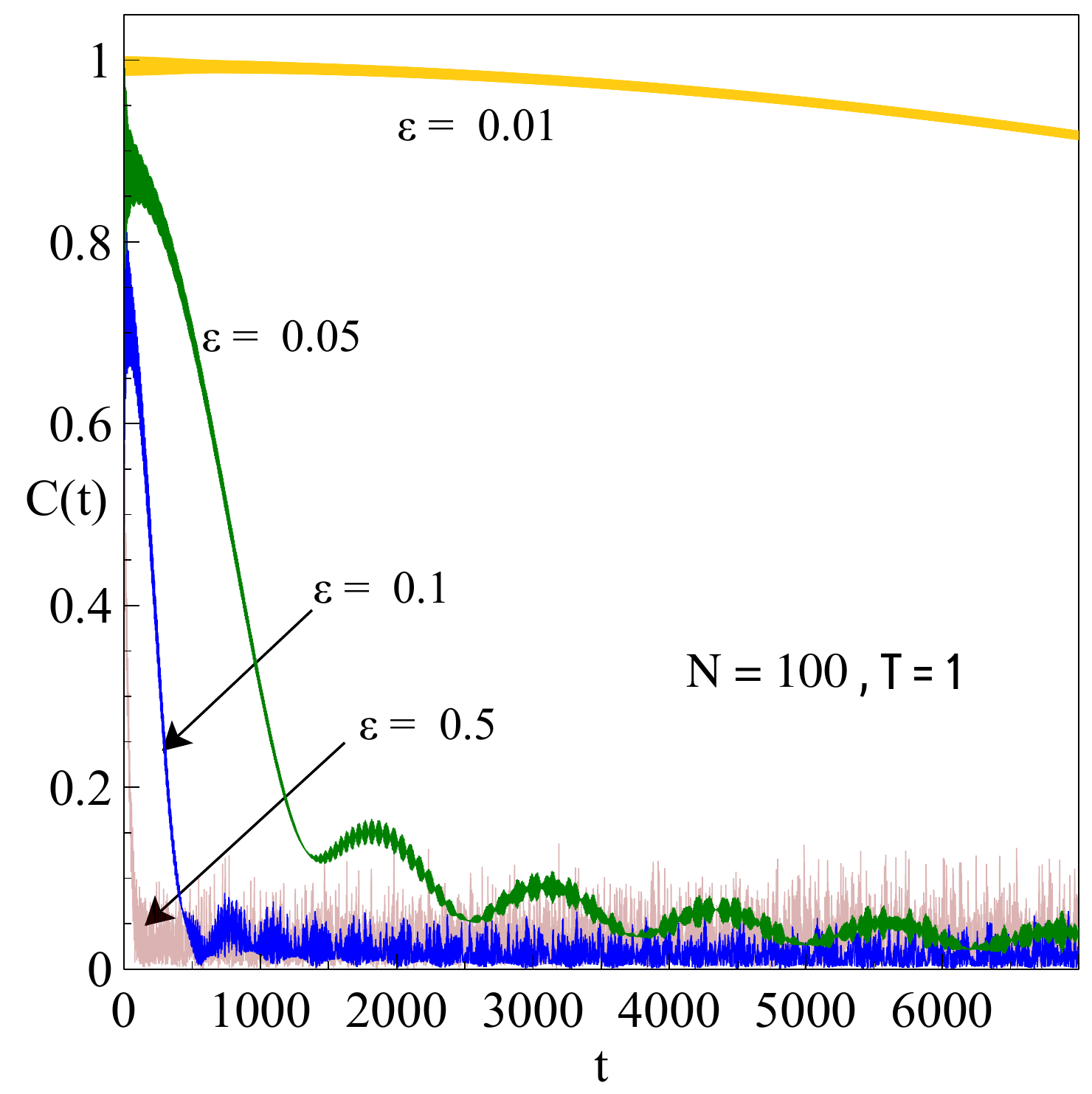}       
    \label{coh-fig2}
}
\subfigure[Variation of coherence $C(t)$ with time $t$ for different number of bath spins.]{
    \includegraphics[width=0.3\textwidth, keepaspectratio]{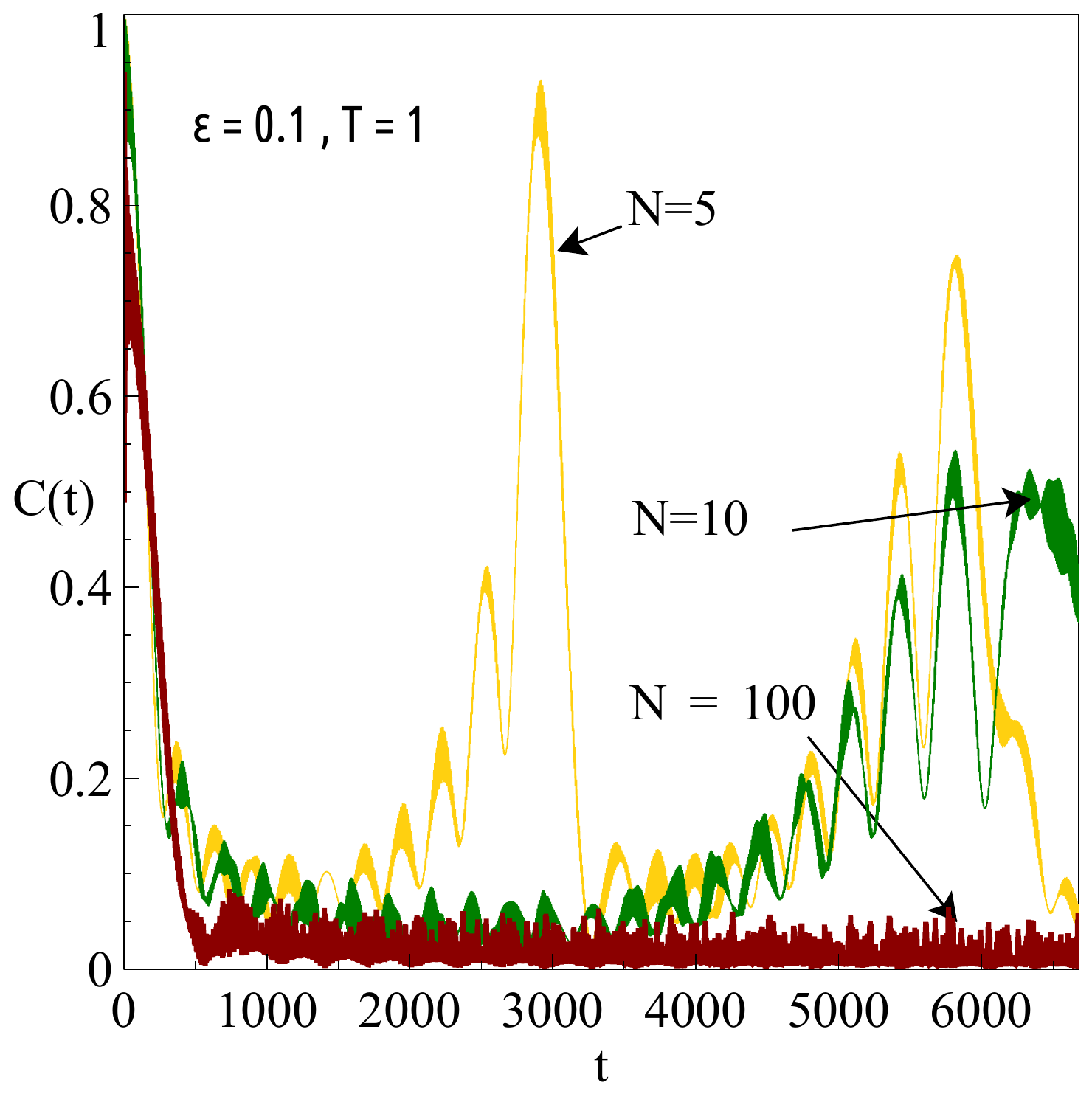}    
    \label{coh-fig3}
}
\subfigure[Variation of entanglement $E(t)$ with time $t$ for different bath temperature.]{
    \includegraphics[width=0.3\textwidth, keepaspectratio]{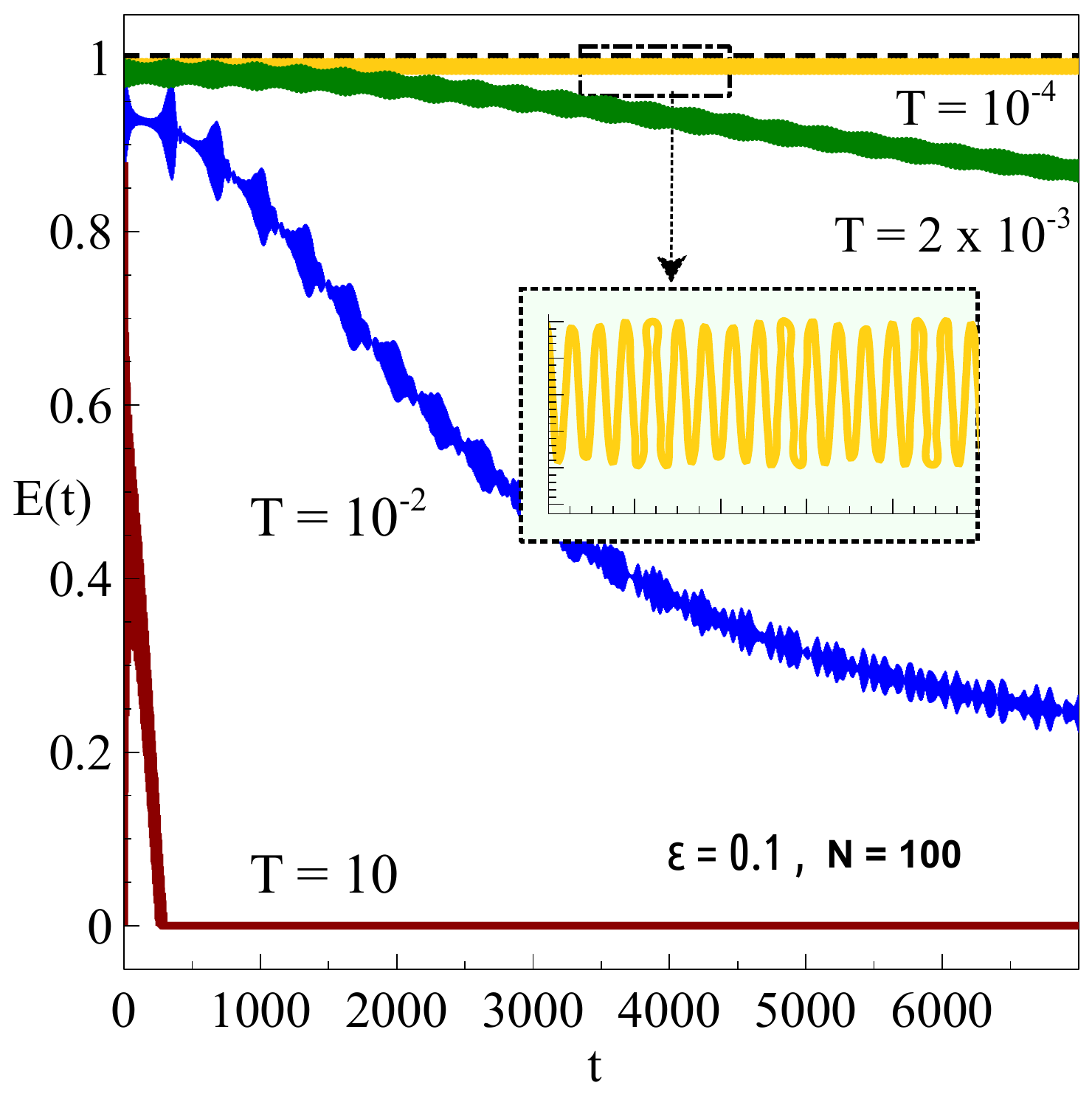}    
    \label{ent-fig1}
}
\subfigure[Variation of entanglement $E(t)$ with time $t$ for different system-bath interaction strength.]{
    \includegraphics[width=0.3\textwidth, keepaspectratio]{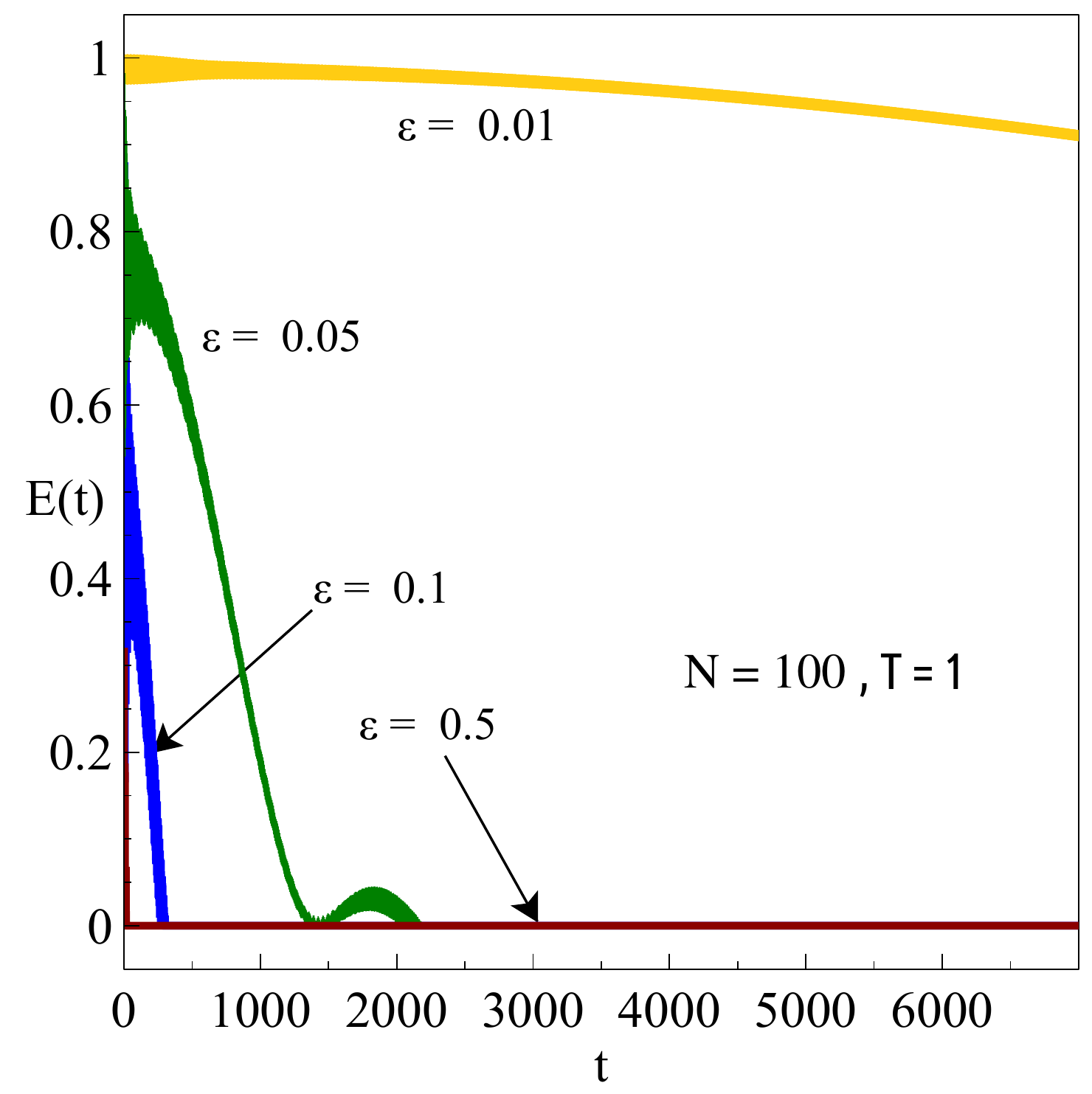}
    \label{ent-fig2}
}
\subfigure[Variation of entanglement $E(t)$ with time $t$ for different number of bath spins.]{
    \includegraphics[width=0.3\textwidth, keepaspectratio]{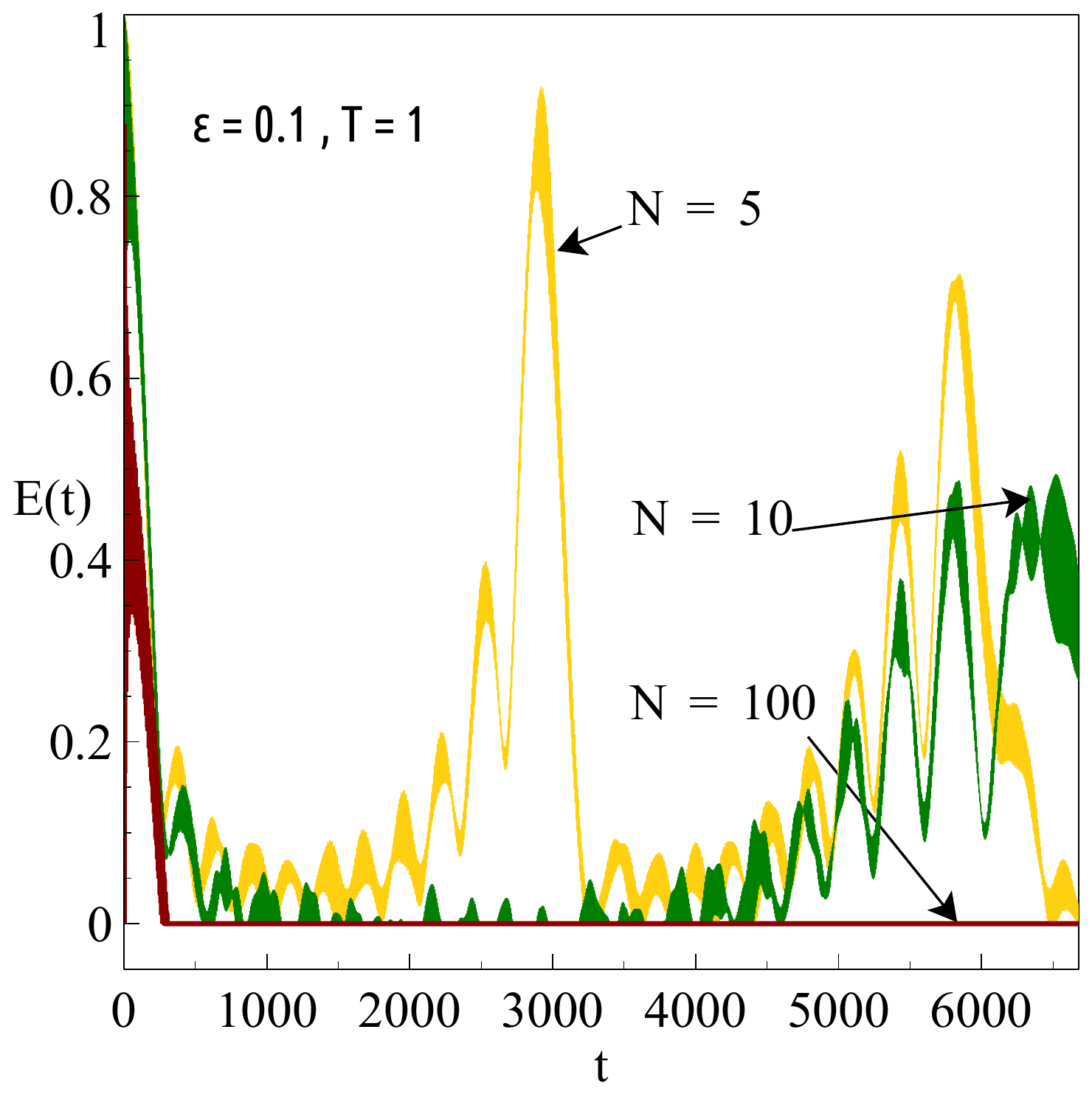}
    \label{ent-fig3}
}
\caption{(Colour online) Dynamics of quantum coherence and entanglement for the central qubit immeresed in the spin bath.}
\label{fig:quantum_dynamics}
\end{figure}

\end{widetext}

We now present the results for time evolution of quantum  coherence and entanglement with the bath temperature $T$, the strength of system-bath interaction $\epsilon$ and number of spins ($N$) in the spin bath attached to the central spin.
If the spin bath is in a very high temperature, we expect the thermal noise to swamp signatures of quantumness, which is broadly confirmed in Fig. \ref{coh-fig1}  and \ref{ent-fig1}. However, small fluctuations in quantum coherence continue to occur testifying to the non-Markovianity  of the dynamics. On the contrary, for low bath temperature, as demonstrated in Fig. \ref{coh-fig1}, quantum coherence does not decay noticeably and for the timespan we considered, it does not dip below a certain value that is in itself quite high.  For intermediate temperatures, coherence broadly decays with increasing decay rate as we increase the bath temperature, but along with small fluctuations due to non-Markovianity. The dynamics of entanglement  as shown in Fig. \ref{ent-fig1}, is quite similar to that of coherence.  At the high temperature limit, the difference with dynamics for quantum coherence lies in the fact that entanglement encounters a sudden death and never revives. This is entirely consistent with the usual observation for many physical systems where quantum coherence turns out to be more robust against noise than entanglement \cite{example,frozencoherence1,frozencoherence2}. In the opposite regime, for low enough temperatures, entanglement dynamics is very much similar to that of coherence. 
Another parameter we can tune is the system-bath interaction strength $\epsilon$, which depending upon the species of the central spin as well as the bath spins, may differ. In case the interaction parameter is too small, the system evolves almost independently from the bath and therefore the coherence and entanglement of the system decay quite slowly as shown in  Fig. \ref{coh-fig2} and  \ref{ent-fig2}. In the opposite limit, if the system-bath interaction is comparable to the energy difference of the spin levels of the central spin, we observe a rapid decay in quantum coherence with the presence of usual non-Markovian fluctuations. Whereas, entanglement decays to zero almost immediately with no revival detected in the time span considered in Fig. \ref{ent-fig2}.
Eq. \eqref{sec1g} also allows us to study the dynamics of coherence for varying number of bath spins. If the number of spins in the bath is large, we observe from Fig. \ref{coh-fig3}, that the coherence rapidly decays and only small fluctuations are subsequently detected. In case the number of spins in the bath is not very large, the evolution of coherence undergoes periodic revivals. The magnitude of such revivals decreases with increasing bath size, eventually reducing to being indistinguishable with smaller fluctuations for large enough number of spins in the bath. As seen in Fig. \ref{coh-fig3},  revivals themselves occur in periodic packets, magnitudes of which decrease steadily with time. On the other hand, if the number of bath particles is quite large, entanglement decays very quickly to zero. However for smaller number of spins in the bath, the entanglement dynamics depicted in Fig. \ref{ent-fig3} is quite similar to the corresponding dynamics of coherence captured earlier in Fig. \ref{coh-fig3}.
%
%
%
%
%

\section{Analysis of time averaged dynamical map}
\label{IV}
In this section we probe the behaviour of long time averaged state of the central spin qubit. We study under what condition the long time averaged state is coherent. We further investigate whether or under what conditions the long time averaged state is a true fixed point of the dynamical map, i.e. independent of initial condition. In connection to that we further study what role the finite size of the environment plays in this context.
The long time averaged state of the central spin qubit is given by 
\beq \label{avgstate_defn}
\overline{\rho} = \lim_{\tau \rightarrow \infty} \frac{\int^{\tau}_{0} \rho (t) dt }{\tau}.
\eeq
Following this definition, we find
\beq \label{Time_Avg_State}
\begin{array}{ll}
\overline{\rho_{11}} = \lim_{\tau \rightarrow \infty} \frac{\int^{\tau}_{0} \rho_{11}(t) dt }{\tau} 
=  \rho_{11} (0) \left(1 - \overline{\alpha} \right)  + \rho_{22}(0) \overline{\beta}, \nonumber \\
\overline{\rho_{12}} = \rho_{12}(0) \overline{\Delta},
\end{array}
\eeq 
where $\overline{\alpha}$, $\overline{\beta}$  and $\overline{\Delta}$ are long time averages of $\alpha(t)$, $\beta(t)$ and $\Delta(t)$ respectively. When we integrate a bounded periodic function over a long time  and divide by the total time elapsed, we can consider the integral being over a large integer number of periods without loss of generality.
Now, 
\beq \label{a2_2}
\begin{array}{ll}
\overline{\alpha} = \sum_{n=0}^N 2(n+1)\epsilon^2\left(1-\frac{n}{2N}\right)\frac{1}{\eta^2}\frac{e^{-\frac{\hbar\omega}{KT}(n/2N-1/2)}}{Z},
\end{array}
\eeq
where the result follows from the fact that average of $sin^{2} (\theta (t)) $ over any integer number of time periods = $\frac{1}{2}$. Similarly we get 
\beq \label{a2_3}
\overline{\beta}=\sum_{n=0}^N 2n\epsilon^2\left(1-\frac{n-1}{2N}\right)\frac{1}{\eta'^2}\frac{e^{-\frac{\hbar\omega}{KT}(n/2N-1/2)}}{Z}
\eeq

\begin{figure}[ht]
\centering
\subfigure[With initial population $\rho_{11} (0)$. ]{
    \includegraphics[width=0.35\textwidth, keepaspectratio]{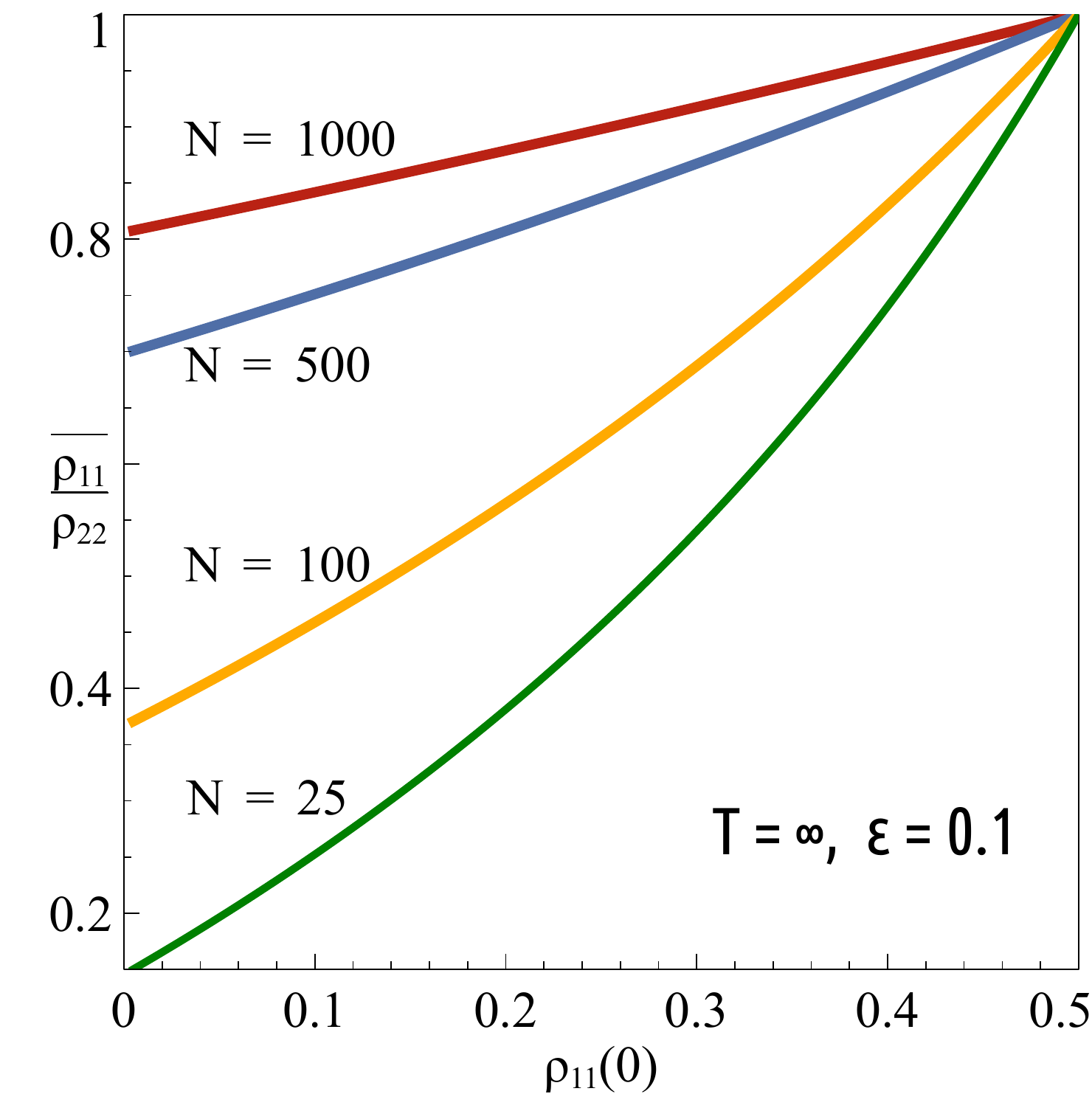}
    \label{steady-fig1}
}
\subfigure[With interaction strength $\epsilon$.]{
    \includegraphics[width=0.35\textwidth, keepaspectratio]{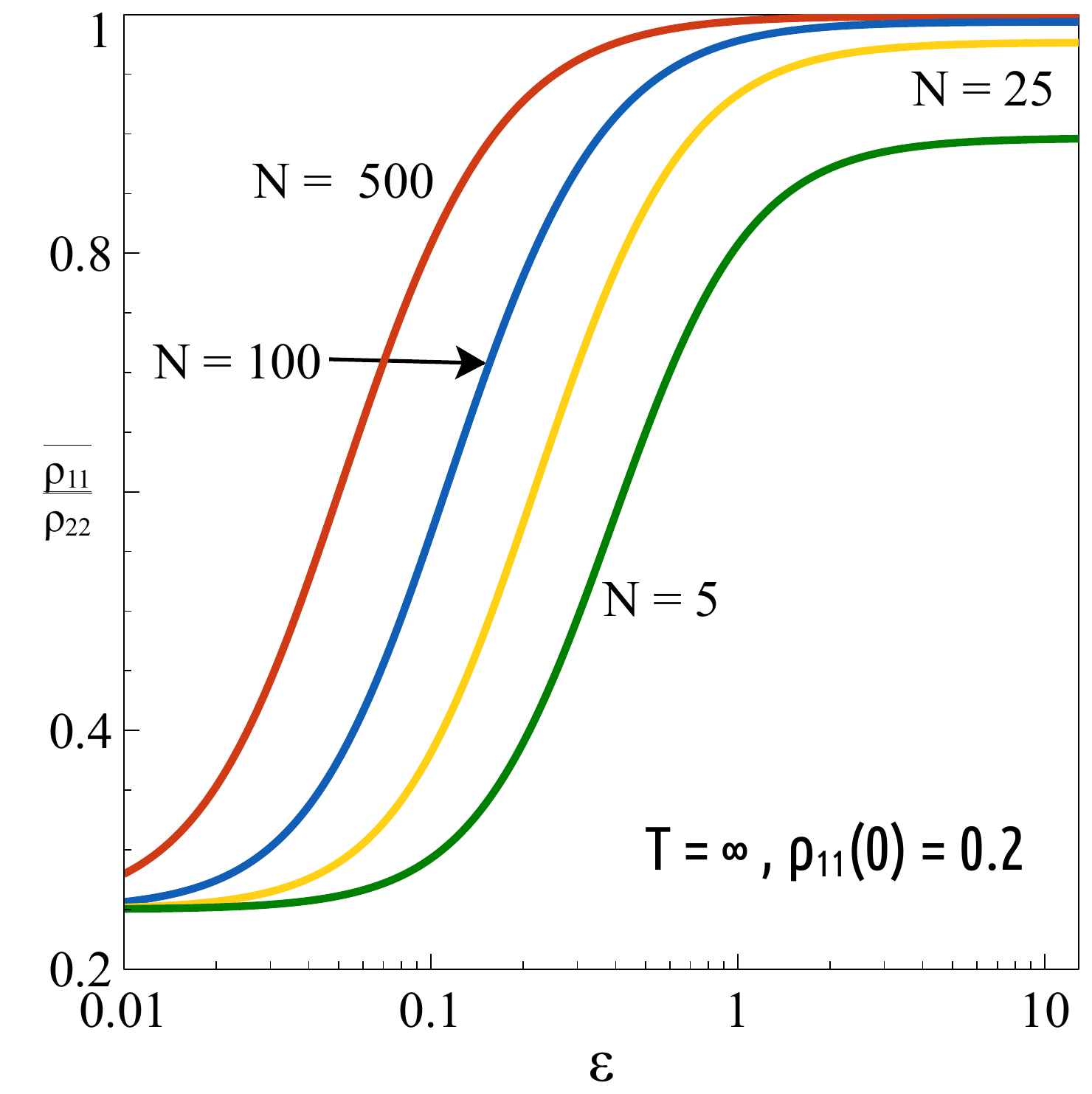}       
    \label{steady-fig2}
}
\caption{(Colour online) Variation of the ratio of long time averaged populations at excited and ground state $\bar{\rho}_{11}/\bar{\rho}_{22}$ with (a) initial population of the excited state $\rho_{11}(0)$ and (b) interaction strength $\epsilon$, keeping the number of bath spins $N$ as a parameter. }
\label{fig:trap_interac}
\end{figure}

The equation for population dynamics shows Eq \eqref{Time_Avg_State} that even the very long time averaged state retains the memory of the initial state, which is a signature of the system being strongly non-Markovian. This initial state dependence is captured in Fig. \ref{steady-fig1}. It is observed that the parameter $\overline{\left(\frac{\rho_{11} }{\rho_{22}}\right)}$ which captures the population distribution for long time averaged state is heavily dependent on the initial ground state population. If the initial population of the ground state increases, so does the population of the ground state for long time averaged state. However, in case the bath is very large, the population statistics for the long time averaged state is markedly less sensitive to the initial population. This leads us to posit that the only true fixed point independent of the initial conditions for this system exists only in the limit $N \rightarrow \infty $. We also observe that in the limit $\rho_{11} (0) = \rho_{22} (0) = \frac{1}{2} $, $\overline{\left(\frac{\rho_{11} }{\rho_{22}}\right)}$ tends towards 1 regardless of bath size $N$ indicating the dynamics is almost unital. Also we should mention that in the thermodynamic limit ($N\rightarrow \infty$), when the temperature of the bath is infinite, the state $\bar{\rho}_{11}=\bar{\rho}_{22}=1/2$ is not only the fixed point of the dynamics but the canonical equilibrium state also. Thus we can conclude that in the limit $N\rightarrow\infty$ and $T\rightarrow\infty$, the present open system dynamics is ergodic. Moreover, we see that the system-bath coupling strength not only affects the timescale of evolution but also plays significant role in the population statistics of the time averaged state. This we can see from  Eq.s \eqref{a2_2} and \eqref{a2_3}, which is also depicted in Fig. \ref{steady-fig2}. Also for most of the cases, we have $\overline{\Delta}=0$. It is interesting to note that the long-time averaged state $\overline{\rho}$ is incoherent in general. This implies, even though quantum coherence or entanglement persists for quite a long time if the bath temperature is very low, as depicted in Fig. \ref{coh-fig1} or Fig. \ref{ent-fig1} respectively, they must eventually decay. It is important to mention that there are specific resonance conditions under which $\overline{\Delta}$ can have finite value, which will be analysed in the following section. 
\subsection{Resonance Condition for long lived quantum coherence}
We have mentioned previously that the long time averaged state is in general diagonal, but for very specific choices of parameter values, this is not true and there indeed is long lived quantum coherence even in the long time averaged state. This can be of significant interest for theoretical and experimental purposes. For the off-diagonal component, the real and imaginary parts of $ \Delta(t)  $, defined as $\Delta_{R}(t)$ and $\Delta_{I}(t)$ respectively equals to
\beq \label{delta}
\begin{array}{ll}
\Delta_{R}(t) = \\
\sum_{n} \cos \frac{\omega t}{2N}\left[  \cos \frac{\eta t}{2} \cos \frac{\eta' t}{2}  + \frac{\left( \omega_{0} - \frac{\omega}{2N} \right) ^2}{\eta \eta'}\sin \frac{\eta t}{2} \sin \frac{\eta' t}{2}\right]\frac{e^{-\frac{\hbar\omega}{KT}(n/2N-1/2)}}{Z}\\
+ \sum_{n}\left( \omega_{0} - \frac{\omega}{2N}\right)  \left[\frac{\sin \frac{\omega t}{2N} \cos \frac{\eta t}{2} \sin \frac{\eta' t}{2}}{\eta'} - \frac{\sin \frac{\omega t}{2N} \sin \frac{\eta t}{2} \cos \frac{\eta' t}{2}}{\eta} \right] \frac{e^{-\frac{\hbar\omega}{KT}(n/2N-1/2)}}{Z}, \\
\Delta_{I}(t) =\\
-\sum_{n}\sin \frac{\omega t}{2N}  \left[ \cos \frac{\eta t}{2} \cos \frac{\eta' t}{2}  +  \frac{\left( \omega_{0} - \frac{\omega}{2N} \right) ^2}{\eta \eta'} \sin \frac{\eta t}{2} \sin \frac{\eta' t}{2}\right]\frac{e^{-\frac{\hbar\omega}{KT}(n/2N-1/2)}}{Z}\\
 +\sum_{n} \left( \omega_{0} - \frac{\omega}{2N}\right)\left[ \frac{\cos \frac{\omega t}{2N} \cos \frac{\eta t}{2} \sin \frac{\eta' t}{2}}{\eta'} - \frac{\cos \frac{\omega t}{2N} \sin \frac{\eta t}{2} \cos \frac{\eta' t}{2}}{\eta} \right] \frac{e^{-\frac{\hbar\omega}{KT}(n/2N-1/2)}}{Z}.
\end{array}
\eeq
We always have 
$$\overline{\sin \theta_{1}(t) \sin \theta_{2} (t) \sin \theta_{3}(t) }  = \overline{\sin \theta_{1}(t) \cos \theta_{2}(t)  \cos \theta_{3}(t)} = 0. $$ For each of the rest of the terms, it can be shown that the criteria for non-zero time averaged coherence reads  \[\frac{\omega}{2N} = \left|\frac{\eta \pm \eta'}{2} \right| . \] For the condition $\frac{\omega}{2N} = \left| \frac{\eta + \eta'}{2} \right|$ to hold, it is easily shown that 
\beq \label{plus_condition}
 N \leq \frac{  \omega}{\omega_{0}}.
\eeq
This, given that $\omega$ and $\omega_{0}$ are usually of the same order of magntitude,  we feel is a rather unrealistic demand on N, since we are concerned with a heat bath, albeit finite sized. We thus concentrate on the other condition $\frac{\omega}{2N} = \left( \frac{\eta  - \eta'}{2} \right)$. The equation $\frac{\omega}{2N} = \left( \frac{\eta  - \eta'}{2} \right) $ can be explicitly expanded out and the following quadratic equation in $n$ is  obtained 
\beq \label{quadratic_in_n}
\begin{array}{ll}
\left( \frac{\epsilon^{4}}{N^{2}} + \frac{ \epsilon^{2} \omega^{2}}{2 N^{3}}\right) n^{2} - \left(\frac{2 \epsilon^{4}}{N} + \frac{  \epsilon^{2} \omega^{2}}{N^{2}}\right) n + \\ \left(\frac{ \omega_{0} \omega^{3} }{4 N^{3}} - \frac{\omega^{2} \omega_{0}^{2}}{4 N^{2}} - \frac{ \epsilon^{2} \omega^{2} }{2 N^{2}} + \epsilon^{4} \right) = 0.
\end{array}
\eeq 
By solving this quadratic equation and noting that the value of $n$ must be an integer, we reach the following equation, which is the resonance condition.
\beq \label{resonance_condition}
N \pm \frac{\epsilon \omega}{2} \frac{\sqrt{\frac{q_1}{8 N^{3}} + \frac{q_2}{16 N^{4}} + \frac{q_3}{32 N^{5}} - \frac{q_4}{64 N^{6}} }}{\frac{\epsilon^{4}}{4 N^{2}} + \frac{\epsilon^{2} \omega^{2}}{8 N^{3}}} \in \mathbb{Z}_{+},
\eeq
with 
$$
\begin{array}{ll}
q_1=\epsilon^{4},~~
q_2=\left( \epsilon^{2} \omega^{2} + \epsilon^{2} \omega_{0}^{2} + 2 \epsilon^{4} \right), \\
q_3=\left( \omega^{2} \omega_{0}^{2} + 2 \epsilon^{2} \omega^{2} - 2 \epsilon^{2} \omega \omega_{0} \right), ~~
q_4=2\omega_{0} \omega^{3},
\end{array}
$$
where $\mathbb{Z}^{+} $ is the set of positive integers $\in [0,N]$. Taking $\omega = \omega_{0} = 1 $ and in the limit $N \gg 1$, we have the resonance condition as
\beq \label{simplified_resonance}
\begin{array}{ll}
 N \pm  \frac{\sqrt{N}}{\epsilon \sqrt{2}} \in \mathbb{Z}^{+},
\end{array}
\eeq 
Thus, if we are interested in obtaining non zero amount of quantum coherence in the long time averaged state, we have to tune the interaction parameter exactly in such a way that $ N \pm  \frac{\sqrt{N}}{\epsilon \sqrt{2}}$ is a positive integer. This is a nice example where precise bath engineering can help us achieve long sustained coherence.
\subsection{Information trapping in the Central Spin System}

Let us now investigate whether or under what condition the dynamical map considered here does have a true fixed point; i.e. the existence of a state which is invariant under the particular dynamics. In order to do that, define the time-averaging map $\overline{\Lambda}$ as the map which takes any initial state $\rho$ to  the corresponding time averaged state $\overline{\rho}$ as given by Eq. \eqref{Time_Avg_State}. Now suppose the system is initially in a state $\rho$. Then a natural question to ask is the following - ``Is the corresponding time averaged state $\overline{\rho}$ invariant under the map $\overline{\Lambda}$ ?" This can only happen when the map $\overline{\Lambda}$ is an idempotent one, i.e. $\overline{\Lambda}^{2} = \overline{\Lambda}$. 
\begin{figure}[htb]
\includegraphics[width=7cm, height=7cm]{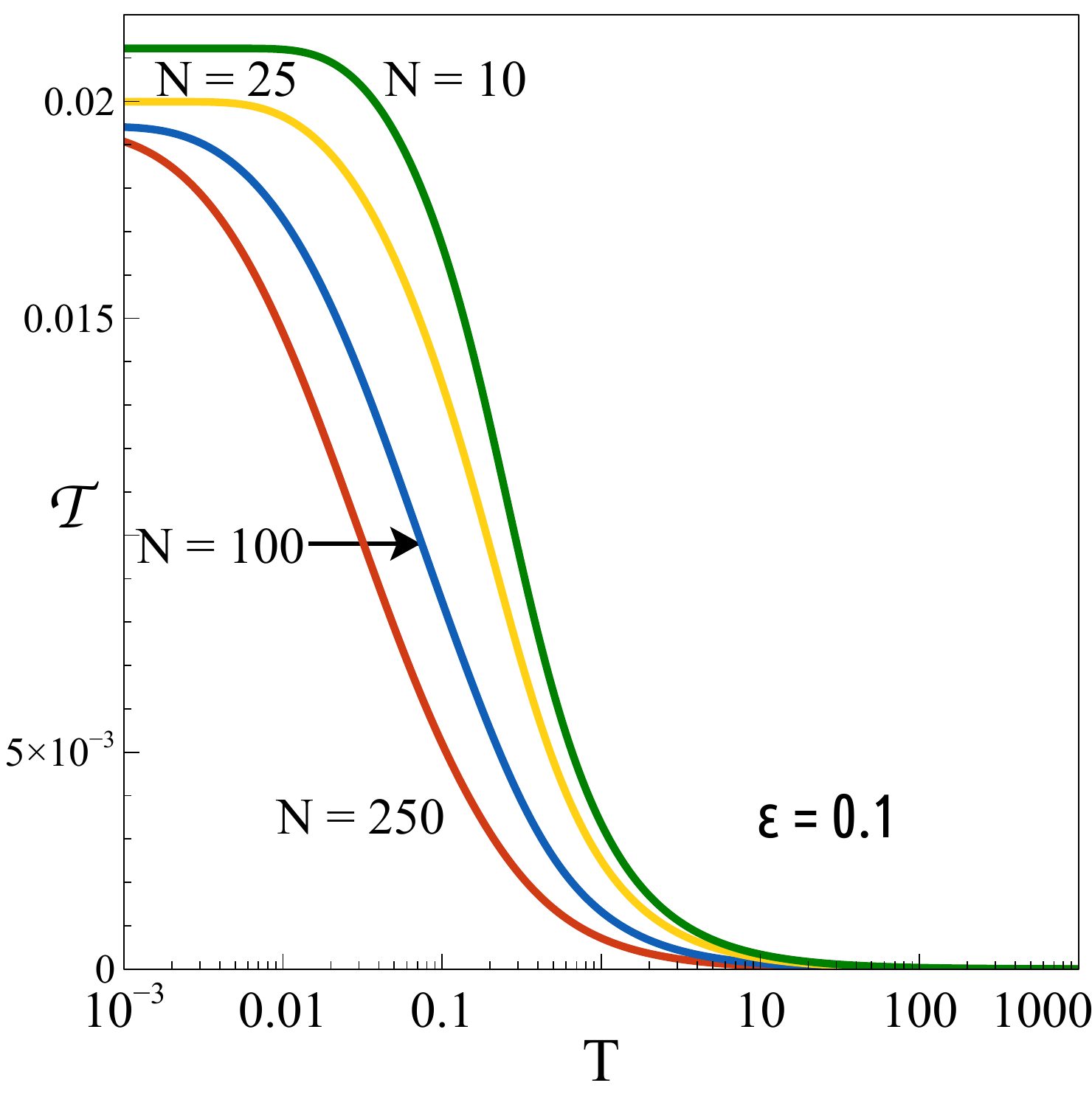}
\caption{(Colour online) Variation of information trapping $\mathcal{T}$ with temperature $T$, keeping the number of bath spins $N$ as a parameter.  }
\label{steady-fig4}
\end{figure}
Clearly, if the time averaged state did not retain the memory of the initial state, this would be the case. Therefore the deviation from idempotence of the map $\overline{\Lambda}$ can serve as a useful measure of the initial state dependence of the system in the long run, which is termed as \textbf{Information Trapping} \cite{smirne} and defined by 
\beq \label{information-trapping}
\mathcal{T} \left( \overline{\Lambda} \right) = \max_{\rho \in \mathcal{H}_{S}} D \left[ \overline{\Lambda}^{2} \rho, \overline{\Lambda} \rho  \right],
\eeq
where D[.,.] is a suitable distance measure on the Hilbert space of the system. Choosing the trace norm as our distance measure, the expression for $\mathcal{T}$ in the central spin model is computed as 
\beq \label{information_trapping_in_our_case}
\mathcal{T} \left( \overline{\Lambda} \right)  = \vert \bar{\beta} - \bar{\alpha} \vert.
\eeq
We immediately note that this quantity vanishes iff $\bar{\beta} = \bar{\alpha}$, which is the case only in the limit $N \rightarrow \infty, T \rightarrow \infty$,  i.e. the thermodynamic and high temperature limit.
The above statement is confirmed in Fig. \ref{steady-fig4}. As we increase the temperature of the bath, the trapped information $\mathcal{T}$ asymptotically vanishes. It is also observed that at any given temperature, the amount of information trapped is greater for a smaller sized bath. This is consistent with the observation that a very large bath is required for $\mathcal{T}$  to vanish.
\begin{figure}[ht]
\centering
\subfigure[Low temperature limit.]{
    \includegraphics[width=0.35\textwidth, keepaspectratio]{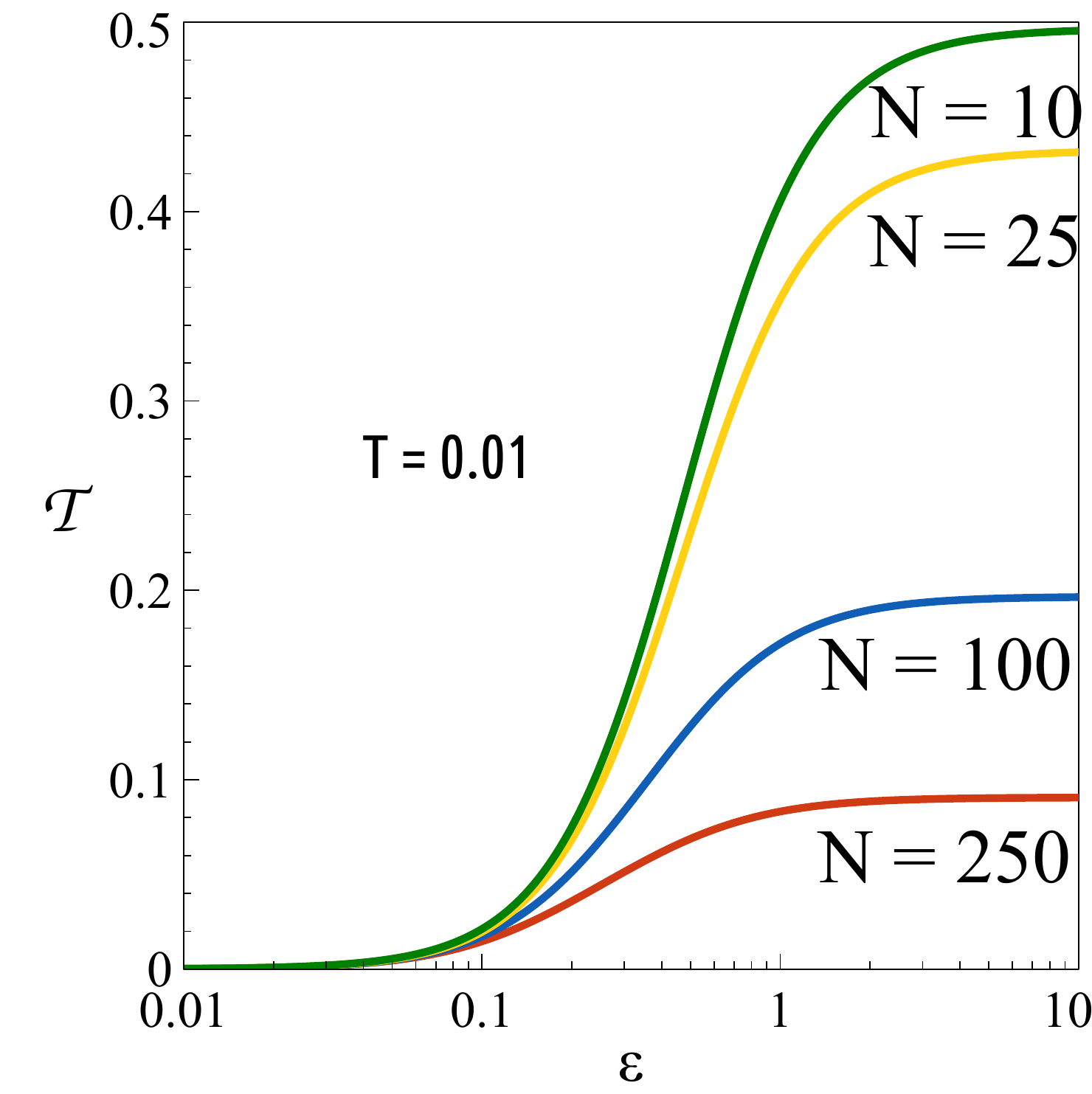}
    \label{steady-fig5}
}
\subfigure[High temperature limit.]{
    \includegraphics[width=0.35\textwidth, keepaspectratio]{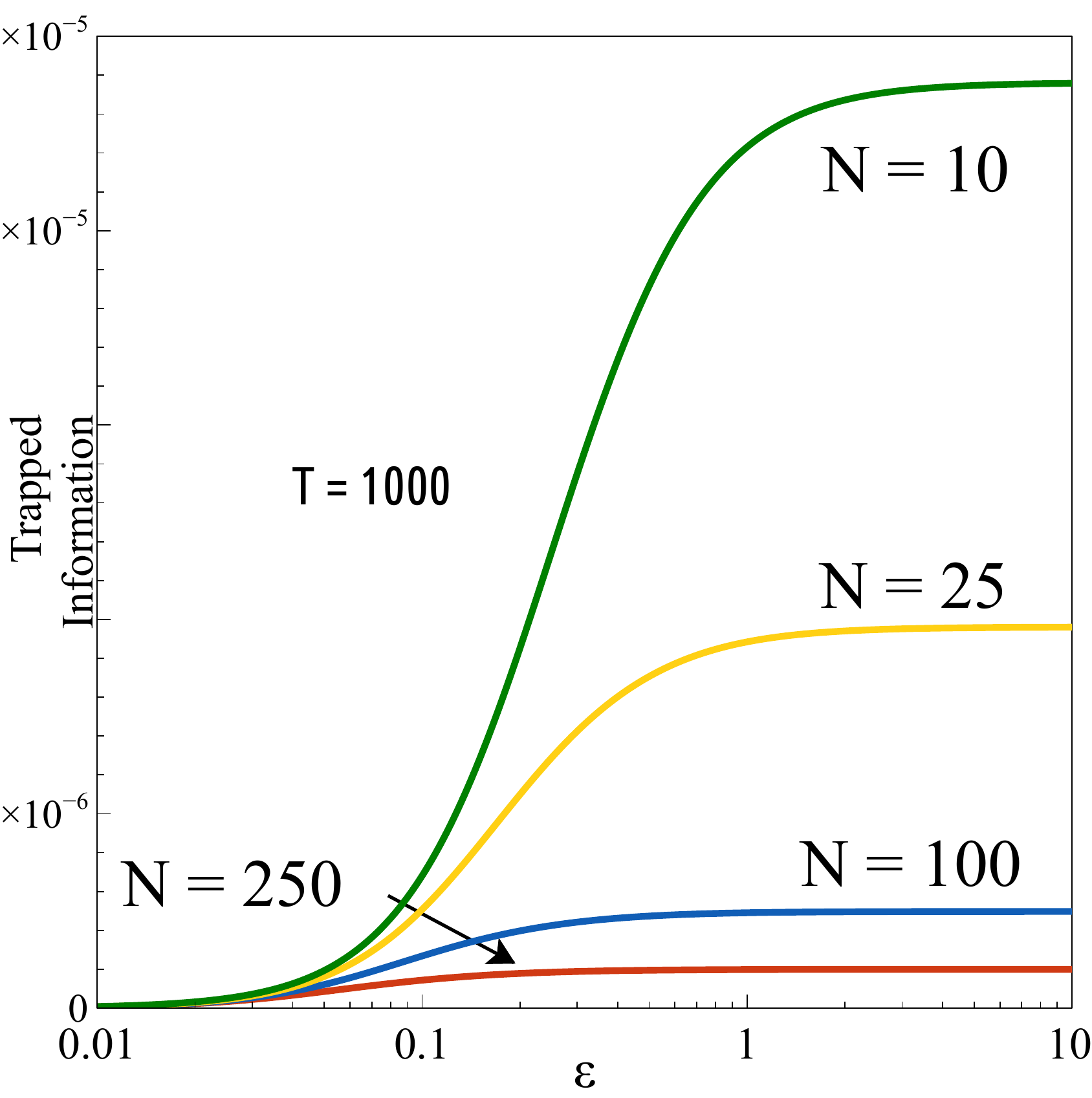}       
    \label{steady-fig6}
}
\caption{(Colour online) Variation of information trapping $\mathcal{T}$ with interaction strength $\epsilon$ at (a) low temperature and (b) high temperature, keeping the number of bath spins $N$ as a parameter. }
\label{fig:trap_interac}
\end{figure}
Fig. \ref{steady-fig5} and \ref{steady-fig6} lead to the observation that as the system-bath coupling gets stronger, the amount of information trapping, i.e. the dependence of the time averaged state on the initial state, also increases. 

\section{Canonical master equation and the process of equilibration} 
\label{III}
Finding the generator of a general dynamical evolution of a quantum system is one of the fundamental problems in the theory of open quantum systems, which leads to a better understanding of the actual nature of decoherence. It is our aim here to derive a canonical master equation without resorting to weak coupling and Born-Markov approximation for the reduced dynamics presented in Eq. \eqref{sec1f}, by virtue of which we will later analyse various thermodynamic aspects of the qubit system. Using the formalism of \citep{hall}, we obtain the following exact time local master equation for the central spin in the Lindblad form.
\beq\label{sec1G}
\begin{array}{ll}
\dot{\rho}(t)=\frac{i}{\hbar}\delta(t)[\rho(t),\sigma_z]+\Gamma_{deph}(t)\left[\sigma_z\rho(t)\sigma_z-\rho(t)\right]\\
~~~~~~+\Gamma_{dis}(t)\left[\sigma_{-}\rho(t)\sigma_{+}-\frac{1}{2}\{\sigma_{+}\sigma_{-},\rho(t)\}\right]\\~~~~~~+\Gamma_{abs}(t)\left[\sigma_{+}\rho(t)\sigma_{-}-\frac{1}{2}\{\sigma_{-}\sigma_{+},\rho(t)\}\right],
\end{array}
\eeq
where $\sigma_{\pm}=\frac{\sigma_x \pm i\sigma_y}{2}$, and $\Gamma_{dis}(t), \Gamma_{abs}(t), \Gamma_{deph}(t)$ are the rates of dissipation, absorption and dephasing processes respectively, and  $\delta(t)$ corresponds to the unitary evolution, respectively, given as
\beq\label{sec1I}
\begin{array}{ll}
\Gamma_{dis}(t)=\left[\frac{d}{dt}\frac{(\alpha(t)-\beta(t))}{2}-\frac{(\alpha(t)-\beta(t)+1)}{2}\frac{d}{dt}\ln(1-\alpha(t)-\beta(t))\right],\\
\\
\Gamma_{abs}(t)=-\left[\frac{d}{dt}\frac{(\alpha(t)-\beta(t))}{2}-\frac{(\alpha(t)-\beta(t)-1)}{2}\frac{d}{dt}\ln(1-\alpha(t)-\beta(t))\right],\\
\\
\Gamma_{deph}(t)=\frac{1}{4}\frac{d}{dt}\left[\ln\left(\frac{1-\alpha(t)-\beta(t)}{|\Delta(t)|^2}\right)\right],\\
\\
\delta(t)=-\frac{1}{2}\frac{d}{dt}\left[\ln\left(1+\left(\frac{\Delta_R(t)}{\Delta_I(t)}\right)^2\right)\right].
\end{array}
\eeq
For the detailed derivation of the master equation, one can look into the Refs. \citep{samyadeb,hall}. Note that the system environment interaction generates a time dependent Hamiltonian evolution in the form of $\delta(t)$. This is analogous to the Lamb-shift correction in the unitary part of the evolution.
Complete positivity \cite{breuer-review,huelga-review,akr1,akr2, laine,rivas} is one of the important properties of a general quantum evolution, following the argument that for any valid quantum dynamical map, the positivity must be preserved if the map is acting on a system which is correlated to an ancilla of any possible dimension. For a Lindblad type evolution, this is guaranteed by the condition $\int_0^t \Gamma_i(s)ds \geq 0$ \cite{kossa}, which can be easily verified for the specific decay rates given in (\ref{sec1I}). However since the dynamical map here is derived starting from an initial product state, complete positivity is always guaranteed \cite{alicki,pechukas}. 
 \subsection{The principle of detailed balance}
\begin{figure}[htb]
\includegraphics[width=7cm, height=6cm]{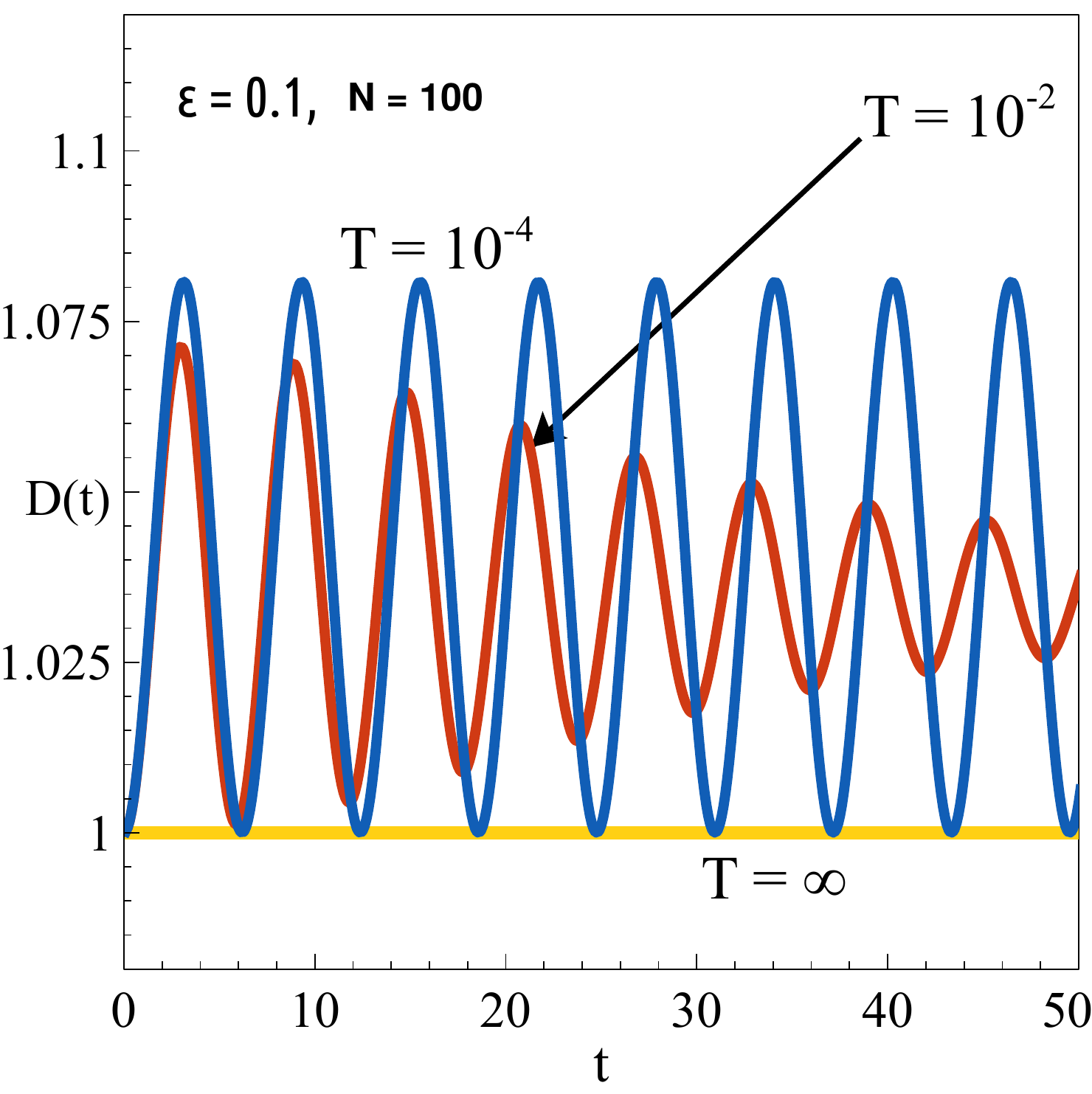}
\caption{(Colour online) Variation of $D(t)$ with time, keeping temperature $T$ as a parameter. $\rho_{11}(0)=0.5$, $\rho_{12}(0)=0$. }
\label{detail-fig1}
\end{figure}

\begin{figure}[htb]
\includegraphics[width=7cm, height=6cm]{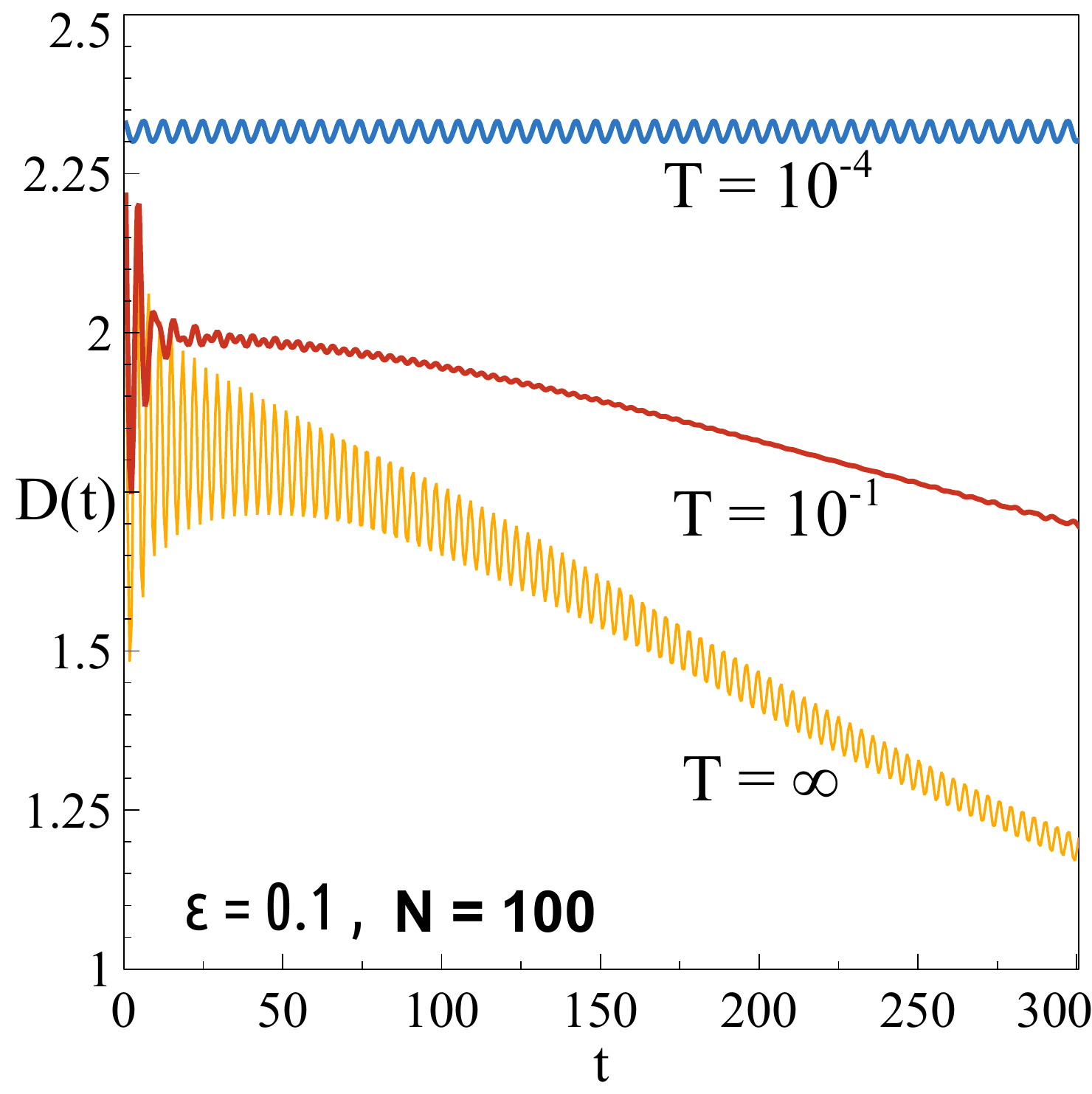}
\caption{(Colour online) Variation of $D(t)$ with time, keeping temperature $T$ as a parameter. $\rho_{11}(0)=0.5$ , $\rho_{12}(0)=0.2$. }
\label{detail-fig1a}
\end{figure}

Here we investigate the process of approach towards steady state for the open system dynamics considered in this paper. There are various different approaches to explore the process of equilibration in an open system dynamics, each of which has their own merit \citep{gogolineisert}. In this work we carry out this investigation for the specific system considered here from a few different aspects, one of which is the quantum detailed balance first introduced by Boltzmann, who used it to prove the famous H-theorem \citep{landau10}. When two or more irreversible processes occur simultaneously, they naturally interfere with each other. If due to the interplay between those different processes, over a sufficient period of evolution time, a certain balance condition between them is reached, then the system reaches a steady state. Consider the Pauli master equation for the atom undergoing such processes \citep{breuerbook} given by 
\beq\label{detailA}
\dot{P}_n=\sum_m \gamma_{nm}P_m -\sum_m \gamma_{mn}P_n,
\eeq
where $P_n$ is the diagonal matrix element of the density operator and $\gamma_{mn}$ is the transition probability for the process $|m\ket\rightarrow |n\ket$. The well known detailed balance condition \cite{kubo,ms} for Pauli master equation is given as $\gamma_{mn}P_n^{(s)}=\gamma_{nm}P_m^{(s)}$, where $P_n^{(s)}$ is diagonal density matrix element at the steady state. We first derive a rate equation of the form of Eq.\eqref{detailA} from the master equation \eqref{sec1G} in order to study the detailed balance for our particular system \citep{hatano, esposito}. Let us consider the unitary matrix $U(t)$, which diagonalizes the system density matrix ($\rho(t)$) as $\rho_D(t)=U(t)\rho(t)U^{\dagger}(t)$. Then we can straightforwardly derive the equation of motion for the diagonalized density matrix as 
\beq\label{detailB}
\begin{array}{ll}
\dot{\rho}_D(t)=\frac{i}{\hbar}\delta(t)[\rho_D(t),\bar{\sigma}_z(t)]\\
~~~~~~+\Gamma_{deph}(t)\left[\bar{\sigma}_z(t)\rho_D(t)\bar{\sigma}_z(t)-\rho_D(t)\right]\\
~~~~~~+\Gamma_{dis}(t)\left[\bar{\sigma}_{-}(t)\rho_D(t)\bar{\sigma}_{+}(t)-\frac{1}{2}\{\bar{\sigma}_{+}(t)\bar{\sigma}_{-}(t),\rho_D(t)\}\right]\\~~~~~~+\Gamma_{abs}(t)\left[\bar{\sigma}_{+}(t)\rho_D(t)\bar{\sigma}_{-}(t)-\frac{1}{2}\{\bar{\sigma}_{-}(t)\bar{\sigma}_{+}(t),\rho_D(t)\}\right],
\end{array}
\eeq
where $\bar{A}_{j}(t)=U(t)A_jU^{\dagger}(t)$. Considering $P_a(t)=\bra a|\rho_D(t)|a\ket$, we get the rate equation similar to the Pauli equation as 
\beq\label{detailC}
\dot{P}_a(t)=\sum_i\sum_b |\bra a|\bar{A}_i(t)|b\ket |^2 P_b(t)-\sum_i \bra a|\bar{A}_i^{\dagger}(t)\bar{A}_i(t)|a\ket P_a(t),
\eeq
where $\bar{A}_i(t)$s are all the Lindblad operators in the diagonal basis as given in Eq. \eqref{detailB}. For the instantaneous steady state we must have $\dot{P}_a(t)=0$, for all $a$. Thus, we have the detailed balance condition 
\beq\label{detailD}
\frac{\sum_i \Gamma_i(t_s)\bra a|\bar{A}_i^{\dagger}(t_s)\bar{A}_i(t_s)|a\ket P_a(t_s)}{\sum_i\sum_b \Gamma_i(t_s)|\bra a|\bar{A}_i(t_s)|b\ket |^2 P_b(t_s)}=1,
\eeq
where $t_s$ is the time at which the system comes to the steady state. From Eq. \eqref{detailB} and \eqref{detailD}, we arrive at \textcolor{black}{the following condition} 
\beq\label{setailE}
D(t_s)=\frac{\Gamma_{dis}(t_s)P_a(t_s)}{\Gamma_{abs}(t_s)P_b(t_s)}=1,
\eeq
where $P_{a,b}(t)=\frac{1}{2}(1\pm \sqrt{(\rho_{11}(t)-\rho_{22}(t))^2+4|\rho_{12}(t)|^2})$
are the eigenvalues of the system density matrix. Any deviation of $D(t)$ from its steady state value, implies that the system has not attained a steady state at that instant of time. The magnitude of such deviations may be regarded as a measure of how far away the system is from equilibrating. In the following we study the time dynamics of deviations from the detailed balance condition Eq. \eqref{setailE}. \\
From Fig. \ref{detail-fig1}, we observe that the deviations from detailed balance condition are quite persistent in the low temperature limit. In the opposite limit, as we go on increasing the bath temperature, Fig. \ref{detail-fig1} shows that the fluctuations in deviation from the detailed balance condition increasingly tend to damp down. In the limit of a completely unpolarized bath, the detailed balance condition is met if the system size is large enough. 
For an initially coherent central qubit, any study of approach towards steady state has to also take the coherence dynamics into account. In the very low temperature limit, the value of quantum coherence (Fig. \ref{coh-fig1}) is encapsulated within a narrow band whose width does not decay much over time. The persistence of coherence in this case implies the deviations are further away from $D(t)=1 $ than in Fig. \ref{detail-fig1}. In the opposite limit of a high temperature bath, quantum coherence dies down very quickly, as seen in Fig. \ref{coh-fig1}. This explains why, just like Fig. \ref{detail-fig1}, $D(t)$ again approaches 1 in Fig. \ref{detail-fig1a}. In the intermediate regime, as we increase the temperature, the approach towards $D(t) =1$ becomes faster.

\begin{figure}[htb]
\includegraphics[width=7cm, height=6cm]{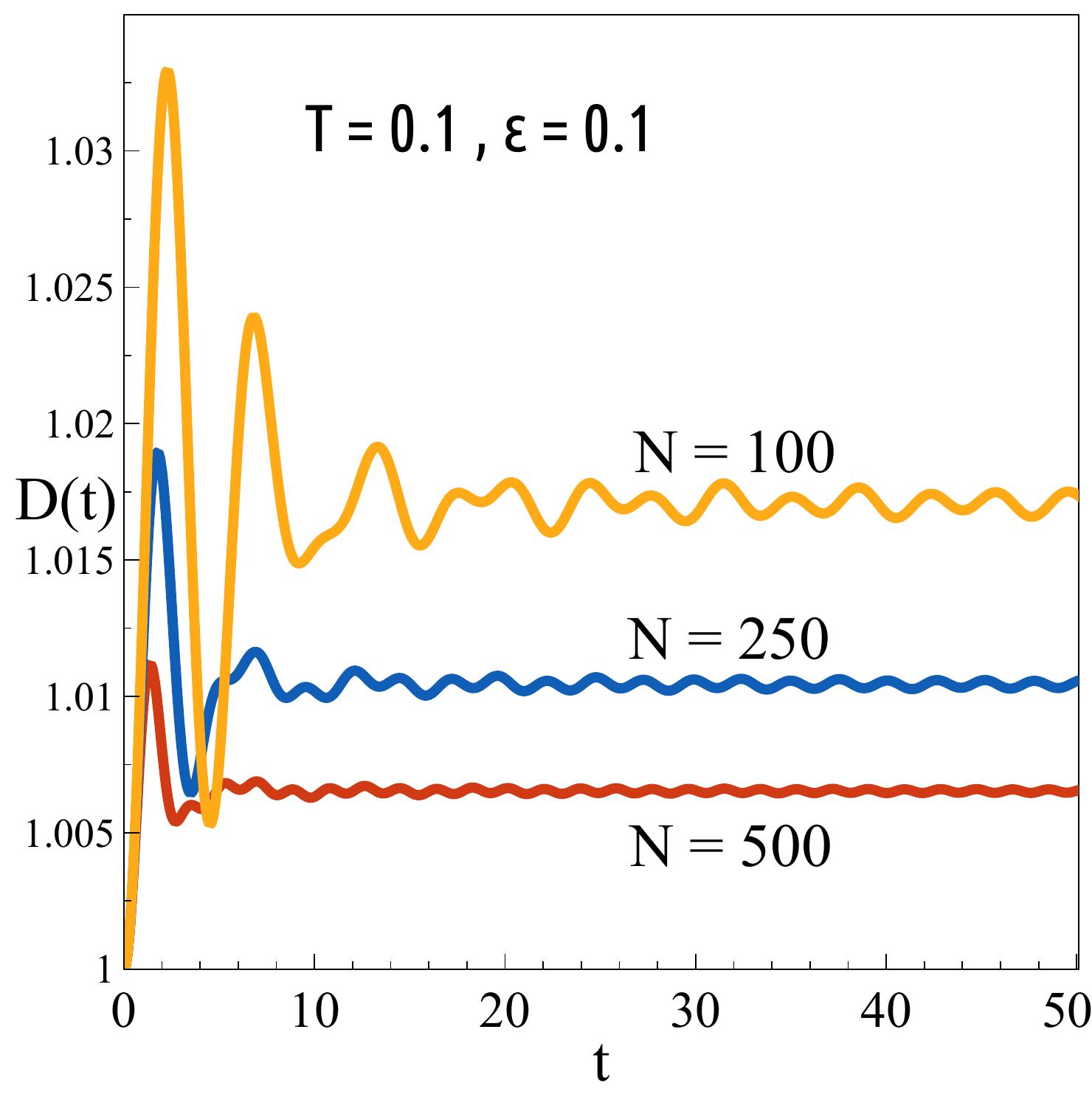}
\caption{(Colour online) Variation of $D(t)$ with time, keeping interaction strength $\epsilon$ as a parameter. $\rho_{11}(0)=0.5, \rho_{12}(0)=0$. }
\label{detail-fig2}
\end{figure}

If the system-bath coupling strength is very weak, we see from Fig. \ref{detail-fig2} that the deviation of $D(t)$ from unity is very small. This is understandable because as the system-bath interaction gets weaker, the change in the state of the system due to the exposure of bath interaction becomes slower and the process becomes more and more quasi-static. Hence, the system remains close to its steady state. As we go on increasing the strength of the interaction, the fluctuations in population levels increase, implying that the deviation from detailed balance condition also increases which is confirmed in Fig. \ref{detail-fig2}. 
\begin{figure}[htb]
\includegraphics[width=7cm, height=6cm]{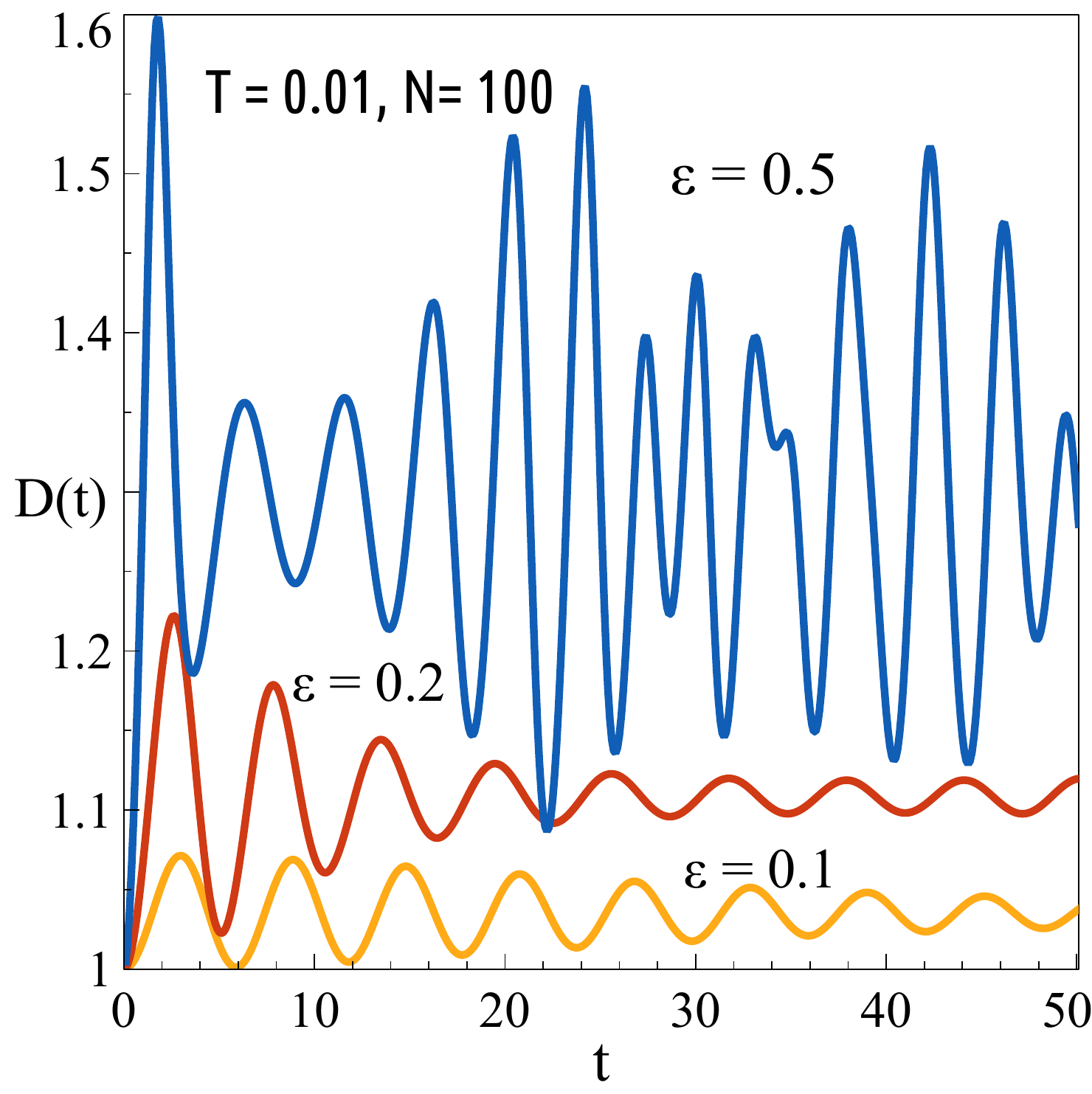}
\caption{(Colour online) Variation of $D(t)$ with time, keeping  $N$ as a parameter. $\rho_{11}(0)=0.5, \rho_{12}(0)=0$.   }
\label{detail-fig3}
\end{figure}
With increasing the bath size, we see from Fig. \ref{detail-fig3} that deviations from detailed balance condition becomes smaller and smaller. This is fully consistent with the observation for many physical systems that  energy exchange and consequent thermalization of a system is  better  facilitated by having a large bath rather than a small ancilla attached to it. 
\subsection{Irreversible Entropy production}
Here we investigate how this system approaches towards a steady state from another thermodynamic perspective, i.e. the phenomenon of irreversible entropy production (IEP). The entropy production rate is formally defined as the negative rate of change of relative entropy between the instantaneous state and the steady state, i.e., $\Sigma(t)=-\frac{d}{dt}S(\rho(t)||\rho_{st})$. For an ideal Markovian evolution, $\Sigma(t)$ is always positive  \cite{spohn}. This happens for few ideal situations and in general is not satisfied. 

\begin{figure}[htb]
\includegraphics[width=7cm, height=6cm]{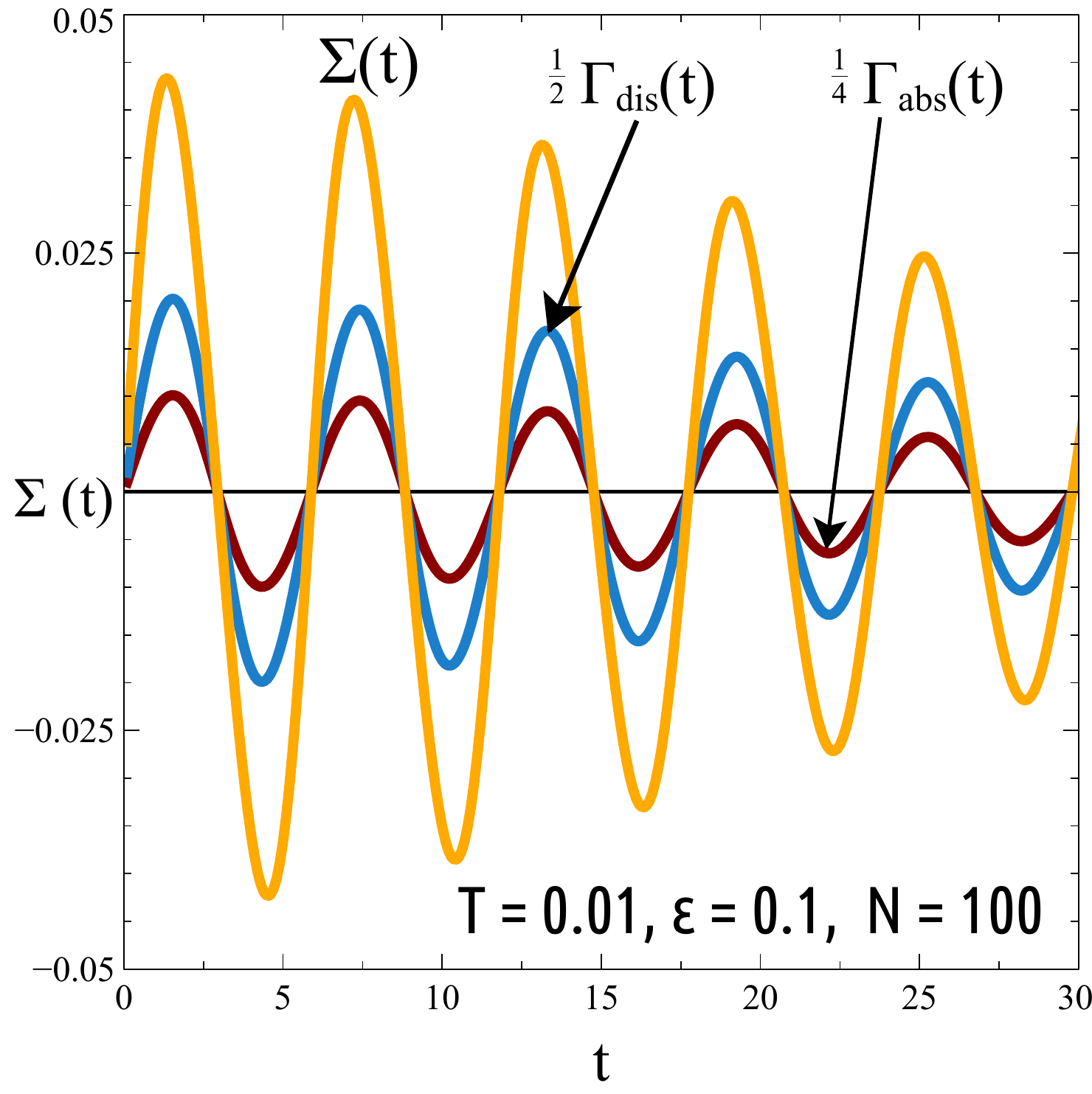}
\caption{(Colour online) Variation of IEP rate $\Sigma (t) $ and Lindblad coefficients for absorption $\Gamma_{abs} (t)$ and dissipation $\Gamma_{dis} (t)$ with time $t$. Initial state $\rho (0)$ is chosen as $\frac{4}{5} | 1\ket\bra 1|$ $+\frac{1}{5}|1\ket\bra 0|+\frac{1}{5}|0\ket\bra 1|+\frac{1}{5}|0\ket\bra 0|$.  }
\label{iep-fig1}
\end{figure}

\begin{figure}[htb]
\includegraphics[width=7cm, height=6cm]{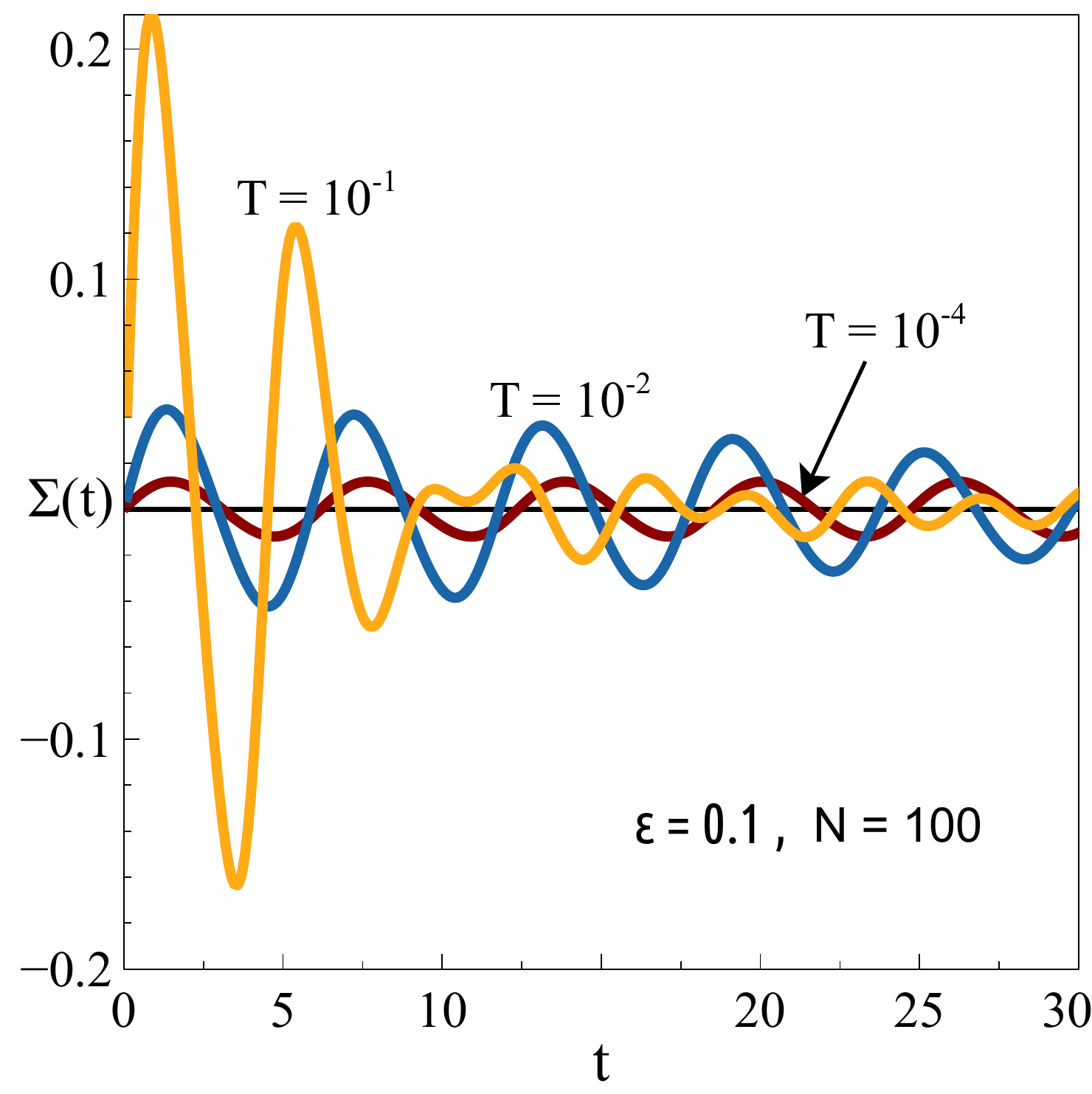}
\caption{(Colour online)  Variation of IEP rate $\Sigma (t) $ with time $t$ for different bath temperatures. Initial state $\rho (0)=\frac{4}{5}|1\ket\bra 1|+\frac{1}{5}|1\ket\bra 0|$ $+\frac{1}{5}|0\ket\bra 1|+\frac{1}{5}|0\ket\bra 0|$. }
\label{iep-fig2}
\end{figure}

\begin{figure}[htb]
\includegraphics[width=7cm, height=6cm]{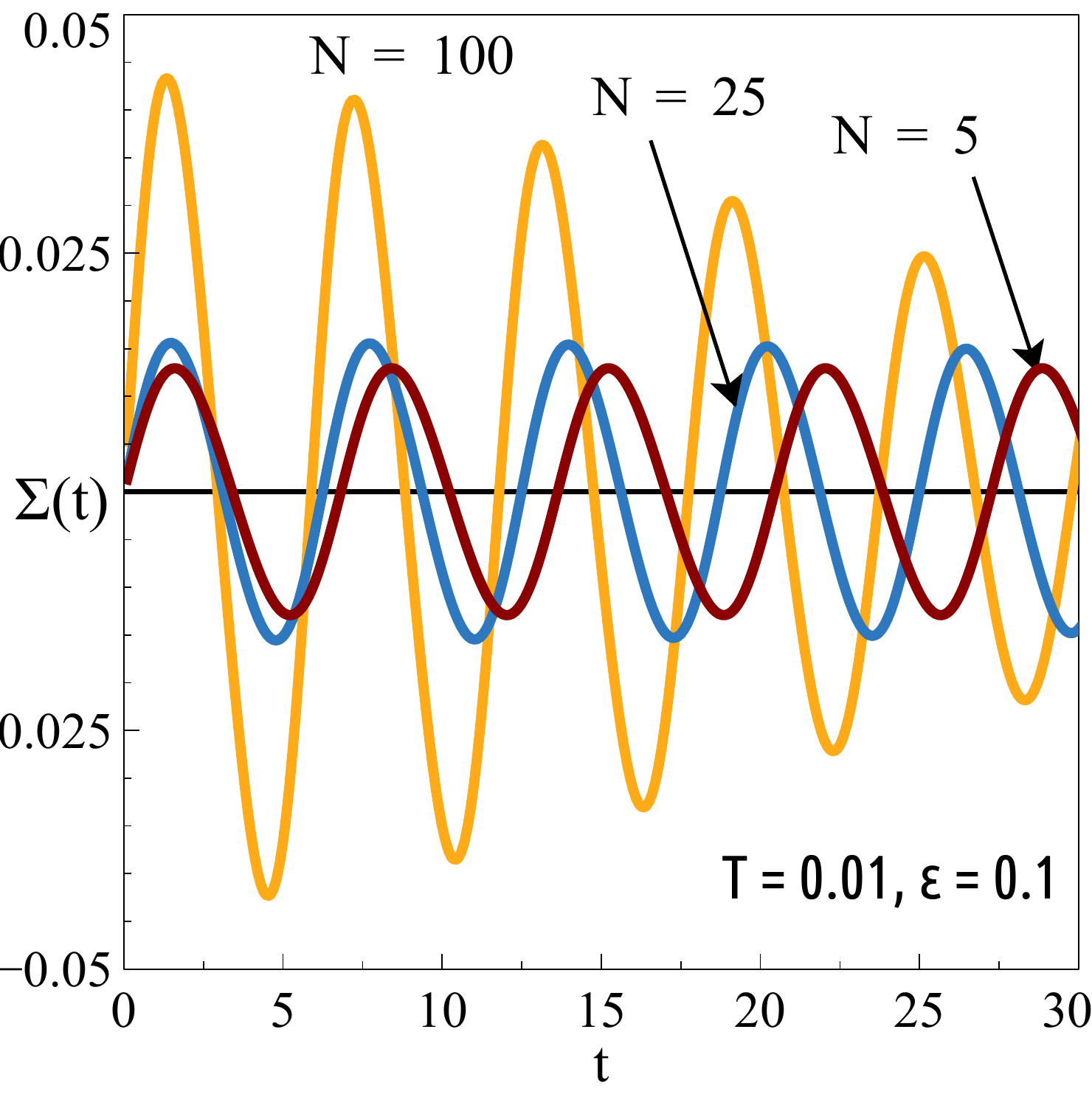}
\caption{(Colour online) Variation of IEP rate $\Sigma (t) $ with time $t$ for different number of bath spins. Initial state $\rho (0) = \frac{4}{5}|1\ket\bra 1|+ $ $\frac{1}{5}|1\ket\bra 0|+ \frac{1}{5}|0\ket\bra 1|+ \frac{1}{5}|0\ket\bra 0|$.  }
\label{iep-fig3}
\end{figure}

The rate equation \eqref{detailC} can be compactly represented as $\dot{P}_a(t)=\sum_b \mathcal{L}_{ab}P_b(t) $, with
$$
\mathcal{L}=\left(\begin{matrix}
-\Gamma_{dis}(t) & \Gamma_{abs}(t) \\
\Gamma_{dis}(t)  & -\Gamma_{abs}(t) 
\end{matrix}\right).
$$
The entropy of the system is defined as $S(t)=-\sum_b P_b(t)\ln P_b(t)$. By differentiating $S(t)$ with respect to time, it can be easily shown that 
\beq\label{entropy1}
\begin{array}{ll}
\dot{S}(t)=\sum_{ab}\mathcal{L}_{ab}P_b(t) \ln\left(\frac{\mathcal{L}_{ab}P_b(t)}{\mathcal{L}_{ba}P_a(t)}\right)
-\sum_{ab}\mathcal{L}_{ab}P_b(t) \ln\left(\frac{\mathcal{L}_{ab}}{\mathcal{L}_{ba}}\right),\\
~~~~~~~=\Sigma(t)+\Phi(t).
\end{array}
\eeq
The first term in the right hand side can be identified as the entropy production rate $\Sigma(t)$ and the second term $\Phi(t)$ defines the effective rate at which entropy is transferred from the environment to the system. For the particular central spin system considered in this paper, the IEP rate is given by
\beq\label{entropy2}
\Sigma(t)=\left(\Gamma_{dis}(t)P_a(t)-\Gamma_{abs}(t)P_b(t)\right)\ln\left(\frac{\Gamma_{dis}(t)P_a(t)}{\Gamma_{abs}(t)P_b(t)}\right).
\eeq
We see from \eqref{entropy2} that IEP rate is related to $D(t)$ and at the time ($t_s$) when system obeys the detailed balance condition,we have $\Sigma(t_s)=0$. We also see from the expression of IEP rate that for Markovian situation (i.e. $\Gamma_{dis}(t),\Gamma_{abs}(t)\geq 0$), it will always be non-negative. This behaviour is illustrated in Fig. \ref{iep-fig1}. Whenever the irreversible entropy production rate $\Sigma (t)$ is negative, the absorption and dissipation rates are also negative and vice versa in the time span we probed. Since negativity of at least one Lindblad coefficient $\Gamma(t)$ is a necessary and sufficient condition \cite{rivas} for non-Markovianity, this leads us to conclude that whenever this system is non-Markovian, a negative IEP rate $\Sigma (t)$ is obtained. While the negativity of IEP rate at any point in the dynamics necessarily implies that the dynamics is non-Markovian, the opposite is not true in general. However, in this illustration we note that the opposite is also true.
If the bath temperature is very low, we have already seen from Fig. \ref{coh-fig1} that the quantum coherence of the central spin qubit  persists for a long time, resulting in persistent deviations from the  steady state detailed balance condition as depicted in Fig. \ref{detail-fig1}. Therefore, it is expected that the IEP rate will also fluctuate and not show any sign of dying down to zero. This is indeed captured in  Fig. \ref{iep-fig2}. In the opposite limit, as we go on increasing the bath temperature, as seen  Fig. \ref{detail-fig1}, the approach towards a steady state becomes quicker. This is again confirmed in Fig. \ref{iep-fig2}, where the fluctuations in IEP rate die down more and more quickly for higher temperatures.
As we have already observed in Fig. \ref{detail-fig3}, the approach towards a steady state through exchange of energy between the system and the bath is quicker for a larger bath. This is again confirmed in Fig. \ref{iep-fig3} which shows the IEP rate becoming smaller and smaller as we increase the bath size. The period of fluctuations also diminish with increasing bath size. 

\section{Conclusion}
\label{V}
In this paper we explore various aspects of a central qubit system in the presence of a  non-interacting thermal spin environment. We solve the Schr\"{o}dinger dynamics of the total state and derive the exact reduced dynamical map for the central qubit. We compute the corresponding Kraus decomposition and evaluate the time evolution of quantum coherence (quantified through the $l_1$ norm) for the qubit in various parameter regions in section \ref{coh_subsection}. We note that as the number of bath spins and the temperature increases, quantum coherence decays steadily with very small fluctuations thus enabling us to conclude that in the thermodynamic limit ($N\rightarrow\infty$) and for sufficiently high temperature, the decay of coherence closely mimics the corresponding behaviour in Markovian systems. We observe quite similar phenomena for quantum entanglement in the same limit, where we see the usual entanglement sudden death. On the contrary, for low temperature both coherence and entanglement, sustain steadily in a band for a very long period of time. This is an important observation having potential practical applications in quantum information processing. For the sake of concretenss, assuming typical order of magnitude values of various parameters governing the dynamics of quantum coherence, we are able to estimate the timescale for which coherence is sustained. Supposing the coupling strength $\epsilon \sim 1$ MHz \citep{prokofiev}, and assuming the spins having intrinsic energies $\sim 100$ MHz \citep{prokofiev}, we can conclude that at room temperature (T = 300 K) and for $N=100$, the value of coherence is guaranteed to be at least 80 percent of the initial coherence for at least $\sim$ 100 $\mu s$. Interestingly, this timescale for guaranteeing at least 80 percent of the initial coherence is not too sensitive on the bath temperature in practice. For example, if we assume the bath to be in a very low temp, say $10^{-4}$ K, then this time increases to only around $\sim 300$ $\mu s$.   It implies that for the open system considered in this paper, the environment can be designed in such particular ways that quantum signatures like coherence or entanglement can be preserved for a  long period of time. For diminishing number of bath spins, steady oscillations of both coherence and entanglement increases both in magnitude and frequency, which can be attributed to the finite size effect. We can contrast the situation with the the extreme case where only one auxillary spin is coupled to the central spin. In that extremal case, the coherence merely oscillates steadily, which is to be expected. But as the number of bath spin increases, the coherence suppression also increases. 
In the second part of our work, we derive the exact canonical master equation for the central qubit, without weak coupling approximation or Born-Markov approximation to study under what condition the central qubit thermalizes with its environment and if not, whether it at all comes to any steady state other than the corresponding thermal state. Probing the quantum detailed balance relation and IEP rate, we conclude that as the completely unpolarized ($T\rightarrow\infty$) spin bath reaches thermodynamic limit, the system equilibrates faster. We see that in the non-Markovian region ($\Gamma_i(t)<0$) of the dynamics, the IEP rate is negative, which is a signature of a system driven away from the equilibrium. However with the increasing number of bath spins and the temperature, we observe that this effect vanishes and the IEP rate remains very close to zero. In fact from further study of long time averaged state and information trapping, we also see that in the mentioned limit, the system actually equilibrates to the corresponding canonical state at infinite temperature. Hence, one may naturally infer that in the limit of $N\rightarrow\infty,T\rightarrow\infty$, the dynamics is ergodic and the bath does not retain the memory of the initial state. However as we deviate from this limit, ergodicity breaks down. In those cases, we observe finite amount of information trapping in the central spin system, which demonstrates that then the bath does hold the memory of the initial state. Perhaps the most important result of the present work is the finding of the existence of coherence in the  long time averaged state of the central spin. We have shown that for specific choices of the system-bath interaction parameter, a resonance condition is satisfied and as a result the long time averaged state retains a finite amount of coherence. Here no external coherent driving is required to preserve this coherence. Our result shows that through precise bath engineering, a spin environment can be manipulated in such a way that it acts as a quantum resource to preserve coherence and potentially entanglement. The presence of such long time quantumness can have potentially far reaching consequence for the construction of quantum thermal machines whose performances are augmented by coherence. 
\section*{Acknowledgements}
SB thanks Anindita Bera of HRI for careful reading of the manuscript. Authors acknowledge financial support from the Department of Atomic Energy, Govt. of India.

\bibliographystyle{apsrev4-1}
\bibliography{spin-bath_finite}

\end{document}